\openin 1 lanlmac
\ifeof 1 
  \message{[Load harvmac]}
  \input harvmac 
\else 
  \message{[Load lanlmac]}
  \input lanlmac
\fi
\closein 1
\input amssym
\input epsf
\noblackbox

\newif\ifhypertex
\ifx\hyperdef\UnDeFiNeD
    \hypertexfalse
    \message{[HYPERTEX MODE OFF]}
    \def\hyperref#1#2#3#4{#4}
    \def\hyperdef#1#2#3#4{#4}
    \def\hypernoname{}
    \def\e@tf@ur#1{}
    
\else
    \hypertextrue
    \message{[HYPERTEX MODE ON]}
    
\fi

\ifx\answ\bigans

\magnification=1200\baselineskip=15pt plus 2pt minus 1pt
%
\advance\voffset by.6truecm
\hsize=6.15truein\vsize=600.truept\hsbody=\hsize\hstitle=\hsize
\else\let\lr=L

\magnification=1000\baselineskip=15pt plus 2pt minus 1pt
%
\hoffset=-0.75truein\voffset=-.0truein
\vsize=6.5truein
\hstitle=8.truein\hsbody=4.75truein
\fullhsize=10truein\hsize=\hsbody
\fi
\parskip=4pt plus 15pt minus 1pt

\newif\iffigureexists
\def\checkex#1{
\relax
\openin 1 #1
   \ifeof 1
      \figureexistsfalse
      \immediate\write20{FIGURE FILE #1 NOT FOUND}
   \else 
      \figureexiststrue
   \fi 
\closein 1}
\def\missbox#1#2{$\vcenter{\hrule
\hbox{\vrule height#1\kern1.truein
\raise.5truein\hbox{#2} \kern1.truein \vrule} \hrule}$}
\def\lfig#1{
\let\labelflag=#1%
\def\numb@rone{#1}%
\ifx\labelflag\UnDeFiNeD%
{\xdef#1{\the\figno}%
\writedef{#1\leftbracket{\the\figno}}%
\global\advance\figno by1%
}\fi{\hyperref{}{figure}{{\numb@rone}}{Fig.~{\numb@rone}}}}
\def\figinsert#1#2#3#4{
\checkex{#4}
\def\figsize{#3}%
\let\flag=#1\ifx\flag\UnDeFiNeD
{\xdef#1{\the\figno}%
\writedef{#1\leftbracket{\the\figno}}%
\global\advance\figno by1%
}\fi
\goodbreak
\midinsert
  \iffigureexists
     \centerline{\epsfysize\figsize\epsfbox{#4}}%
  \else
     \vskip.05truein
     \centerline{\missbox\figsize{#4 not found!}}%
     \vskip.05truein
  \fi
{\smallskip%
\leftskip 4pc \rightskip 4pc%
\noindent\ninepoint\sl \baselineskip=11pt%
{\bf{\hyperdef\hypernoname{figure}{{#1}}{Fig.~{#1}}}:~}#2%
\smallskip}\bigskip\endinsert%
}

\newcount\tabno
\tabno=1
\def\ltab#1{
\let\labelflag=#1%
\def\numb@rone{#1}%
\ifx\labelflag\UnDeFiNeD{%
  \xdef#1{\the\tabno}%
  \writedef{#1\leftbracket{\the\tabno}}%
  \global\advance\tabno by1%
}%
\fi%
{\hyperref{}{table}{{\numb@rone}}{Table~{\numb@rone}}}}
\def\tabinsert#1#2#3{
\let\flag=#1
\ifx\flag\UnDeFiNeD
  {\xdef#1{\the\tabno}
   \writedef{#1\leftbracket{\the\tabno}}
   \global\advance\tabno by1 }
\fi
\vbox{\bigskip #3 \smallskip
\leftskip 4pc \rightskip 4pc
\noindent\ninepoint\sl \baselineskip=11pt
{\bf{\hyperdef\hypernoname{table}{{#1}}{Table~{#1}}}.~}#2
\smallskip}
\bigskip}

\def\ie{{\it i.e.\ }}
\def\eg{{\it e.g.\ }}
\def\cf{{\it cf.\ }}
\def\ss{\scriptstyle}
\def\sss{\scriptscriptstyle}
\def\ssk{\hskip-4pt\sss}
\def\sskk{\hskip-8pt\sss}
\def\skcr{\hskip-4pt\cr}
\def\Ext{{\rm Ext}}
\def\Hom{{\rm Hom}}
\def\hom{{\rm hom}}
\def\Cone{{\rm Cone}}
\def\Diag{{\rm Diag}}
\def\Res{{\rm Res}}
\def\Str{{\rm Str}}
\def\ch{{\rm ch}}
\def\ket#1{\left| #1 \right\rangle}
\def\bra#1{\left\langle #1 \right|}
\def\braket#1#2{\left\langle #1 \vphantom{\ket{#2}} \right|\hskip-3pt
                        \left. \vphantom{\bra{#1}} #2 \right\rangle}
\def\IC{{\Bbb C}}
\def\IP{{\Bbb P}}

\def\IZ{{\Bbb Z}}

\def\cL{{\cal L}}


\lref\AshokZB{
S.~K.~Ashok, E.~Dell'Aquila and D.~E.~Diaconescu,
``Fractional branes in Landau-Ginzburg orbifolds,''
Adv.\ Theor.\ Math.\ Phys.\  {\bf 8}, 461 (2004)
[arXiv:hep-th/0401135].
}

\lref\AspinwallDZ{
P.~S.~Aspinwall and M.~R.~Douglas,
``D-brane stability and monodromy,''
JHEP {\bf 0205}, 031 (2002)
[arXiv:hep-th/0110071].
}

\lref\AspinwallIB{
P.~S.~Aspinwall,
``The Landau-Ginzburg to Calabi-Yau dictionary for D-branes,''
arXiv:hep-th/0610209.
}

\lref\AspinwallJR{
P.~S.~Aspinwall,
``D-branes on Calabi-Yau manifolds,''
arXiv:hep-th/0403166.
}

\lref\AspinwallPU{
P.~S.~Aspinwall and A.~E.~Lawrence,
``Derived categories and zero-brane stability,''
JHEP {\bf 0108}, 004 (2001)
[arXiv:hep-th/0104147].
}

\lref\AshokXQ{
S.~K.~Ashok, E.~Dell'Aquila, D.~E.~Diaconescu and B.~Florea,
``Obstructed D-branes in Landau-Ginzburg orbifolds,''
Adv.\ Theor.\ Math.\ Phys.\  {\bf 8}, 427 (2004)
[arXiv:hep-th/0404167].
}

\lref\BrunnerDC{
I.~Brunner, M.~Herbst, W.~Lerche and B.~Scheuner,
``Landau-Ginzburg realization of open string TFT,''
arXiv:hep-th/0305133.
}

\lref\BrunnerEG{
I.~Brunner and J.~Distler,
``Torsion D-branes in nongeometrical phases,''
Adv.\ Theor.\ Math.\ Phys.\  {\bf 5}, 265 (2002)
[arXiv:hep-th/0102018].
}

\lref\BrunnerJQ{
I.~Brunner, M.~R.~Douglas, A.~E.~Lawrence and C.~R\"omelsberger,
``D-branes on the quintic,''
JHEP {\bf 0008}, 015 (2000)
[arXiv:hep-th/9906200].
}

\lref\BrunnerMT{
I.~Brunner, M.~Herbst, W.~Lerche and J.~Walcher,
``Matrix factorizations and mirror symmetry: The cubic curve,''
arXiv:hep-th/0408243.
}

\lref\BrunnerTC{
I.~Brunner, M.~R.~Gaberdiel and C.~A.~Keller,
``Matrix factorisations and D-branes on K3,''
JHEP {\bf 0606}, 015 (2006)
[arXiv:hep-th/0603196].
}

\lref\BrunnerSK{
I.~Brunner, J.~Distler and R.~Mahajan,
``Return of the torsion D-branes,''
Adv.\ Theor.\ Math.\ Phys.\  {\bf 5}, 311 (2002)
[arXiv:hep-th/0106262].
}

\lref\CandelasRM{
P.~Candelas, X.~C.~De La Ossa, P.~S.~Green and L.~Parkes,
``A pair of Calabi-Yau manifolds as an exactly soluble superconformal
theory,''
Nucl.\ Phys.\ B {\bf 359}, 21 (1991).
}

\lref\DiaconescuVP{
D.~E.~Diaconescu and C.~R\"omelsberger,
``D-branes and bundles on elliptic fibrations,''
Nucl.\ Phys.\ B {\bf 574}, 245 (2000)
[arXiv:hep-th/9910172].
}

\lref\DistlerYM{
J.~Distler, H.~Jockers and H.~J.~Park,
``D-brane monodromies, derived categories and boundary linear sigma
models,''
arXiv:hep-th/0206242.
}

\lref\DouglasFJ{
M.~R.~Douglas,
``Dirichlet branes, homological mirror symmetry, and stability,''
arXiv:math.ag/0207021.
}

\lref\DouglasAH{
M.~R.~Douglas, B.~Fiol and C.~R\"omelsberger,
``Stability and BPS branes,''
JHEP {\bf 0509}, 006 (2005)
[arXiv:hep-th/0002037].
}

\lref\DouglasGI{
M.~R.~Douglas,
``D-branes, categories and N = 1 supersymmetry,''
J.\ Math.\ Phys.\  {\bf 42}, 2818 (2001)
[arXiv:hep-th/0011017].
}

\lref\DouglasVM{
M.~R.~Douglas,
``Topics in D-geometry,''
Class.\ Quant.\ Grav.\  {\bf 17}, 1057 (2000)
[arXiv:hep-th/9910170].
}

\lref\DouglasQW{
M.~R.~Douglas, B.~Fiol and C.~R\"omelsberger,
``The spectrum of BPS branes on a noncompact Calabi-Yau,''
JHEP {\bf 0509}, 057 (2005)
[arXiv:hep-th/0003263].
}

\lref\GepnerVZ{
D.~Gepner,
``Exactly solvable string compactifications on manifolds of SU(N) holonomy,''
Phys.\ Lett.\ B {\bf 199}, 380 (1987).
}

\lref\GepnerQI{
D.~Gepner,
``Space-time supersymmetry in compactified string theory and superconformal
models,"
Nucl.\ Phys.\ B {\bf 296}, 757 (1988).
}

\lref\GovindarajanIM{
S.~Govindarajan, H.~Jockers, W.~Lerche and N.~P.~Warner,
``Tachyon condensation on the elliptic curve,''
arXiv:hep-th/0512208.
}

\lref\GovindarajanKR{
S.~Govindarajan and T.~Jayaraman,
``Boundary fermions, coherent sheaves and D-branes on Calabi-Yau
manifolds,''
Nucl.\ Phys.\ B {\bf 618}, 50 (2001)
[arXiv:hep-th/0104126].
}

\lref\GovindarajanUY{
S.~Govindarajan and H.~Jockers,
``Effective superpotentials for B-branes in Landau-Ginzburg models,''
JHEP {\bf 0610}, 060 (2006)
[arXiv:hep-th/0608027].
}

\lref\GovindarajanVI{
S.~Govindarajan and T.~Jayaraman,
``D-branes, exceptional sheaves and quivers on Calabi-Yau manifolds: From
Mukai to McKay,''
Nucl.\ Phys.\ B {\bf 600}, 457 (2001)
[arXiv:hep-th/0010196].
}

\lref\GreeneTX{
B.~R.~Greene and Y.~Kanter,
``Small volumes in compactified string theory,''
Nucl.\ Phys.\ B {\bf 497}, 127 (1997)
[arXiv:hep-th/9612181].
}

\lref\HerbstWorkP{
M.~Herbst, K.~Hori and D.~Page,
work in progress.
}

\lref\HerbstAX{
M.~Herbst and C.~I.~Lazaroiu,
 ``Localization and traces in open-closed topological Landau-Ginzburg
models,''
JHEP {\bf 0505}, 044 (2005)
[arXiv:hep-th/0404184].
}

\lref\HoriBX{
K.~Hori,
``Boundary RG flows of N = 2 minimal models,''
arXiv:hep-th/0401139.
}

\lref\HoriJA{
K.~Hori and J.~Walcher,
``F-term equations near Gepner points,''
JHEP {\bf 0501}, 008 (2005)
[arXiv:hep-th/0404196].
}

\lref\HoriZD{
K.~Hori and J.~Walcher,
``D-branes from matrix factorizations,''
Comptes Rendus Physique {\bf 5}, 1061 (2004)
[arXiv:hep-th/0409204].
}

\lref\HerbstZM{
M.~Herbst, C.~I.~Lazaroiu and W.~Lerche,
``D-brane effective action and tachyon condensation in topological minimal
models,''
JHEP {\bf 0503}, 078 (2005)
[arXiv:hep-th/0405138].
}

\lref\KapustinBI{
A.~Kapustin and Y.~Li,
``D-branes in Landau-Ginzburg models and algebraic geometry,''
JHEP {\bf 0312}, 005 (2003)
[arXiv:hep-th/0210296].
}

\lref\KapustinGA{
A.~Kapustin and Y.~Li,
``Topological correlators in Landau-Ginzburg models with boundaries,''
Adv.\ Theor.\ Math.\ Phys.\  {\bf 7}, 727 (2004)
[arXiv:hep-th/0305136].
}

\lref\KnappRD{
J.~Knapp and H.~Omer,
``Matrix factorizations, minimal models and Massey products,''
JHEP {\bf 0605}, 064 (2006)
[arXiv:hep-th/0604189].
}

\lref\KontsevichICM{
M.~Kontsevich,
``Homological Algebra of Mirror Symmetry,"
arXiv:alg-geom/9411018.
}

\lref\LazaroiuJM{
C.~I.~Lazaroiu,
``Generalized complexes and string field theory,''
JHEP {\bf 0106}, 052 (2001)
[arXiv:hep-th/0102122].
}

\lref\LazaroiuZI{
C.~I.~Lazaroiu,
``On the boundary coupling of topological Landau-Ginzburg models,''
JHEP {\bf 0505}, 037 (2005)
[arXiv:hep-th/0312286].
}

\lref\LazaroiuWM{
C.~I.~Lazaroiu,
``Graded D-branes and skew-categories,''
arXiv:hep-th/0612041.
}

\lref\LercheCS{
W.~Lerche, D.~L\"ust and N.~P.~Warner,
``Duality symmetries in N=2 Landau-Ginzburg models,''
Phys.\ Lett.\ B {\bf 231}, 417 (1989).
}

\lref\MayrAS{
P.~Mayr,
``Phases of supersymmetric D-branes on K\"ahler manifolds and the McKay
correspondence,''
JHEP {\bf 0101}, 018 (2001)
[arXiv:hep-th/0010223].
}

\lref\OrlovYP{
D.~Orlov,
``Triangulated categories of singularities and D-branes in  Landau-Ginzburg
models,''
arXiv:math.ag/0302304.
}

\lref\OrlovMAG{
D.~Orlov,
``Derived categories of coherent sheaves and triangulated categories of
singularities,''
arXiv:math.ag/0503632.
}

\lref\PolchinskiSM{
J.~Polchinski and A.~Strominger,
``New Vacua for Type II String Theory,''
Phys.\ Lett.\ B {\bf 388}, 736 (1996)
[arXiv:hep-th/9510227].
}

\lref\PolishchukDB{
A.~Polishchuk and E.~Zaslow,
``Categorical Mirror Symmetry: The Elliptic Curve,''
Adv.\ Theor.\ Math.\ Phys.\  {\bf 2}, 443 (1998)
[arXiv:math.ag/9801119].
}

\lref\RecknagelSB{
A.~Recknagel and V.~Schomerus,
``D-branes in Gepner models,''
Nucl.\ Phys.\ B {\bf 531}, 185 (1998)
[arXiv:hep-th/9712186].
}

\lref\SeidelIA{
P.~Seidel and R.~P.~Thomas,
``Braid group actions on derived categories of coherent sheaves,''
arXiv:math.ag/0001043.
}

\lref\WalcherTX{
J.~Walcher,
``Stability of Landau-Ginzburg branes,''
J.\ Math.\ Phys.\  {\bf 46}, 082305 (2005)
[arXiv:hep-th/0412274].
}

\lref\WittenYC{
E.~Witten,
``Phases of N = 2 theories in two dimensions,''
Nucl.\ Phys.\ B {\bf 403}, 159 (1993)
[arXiv:hep-th/9301042].
}


\lref\BKGriffithsHarris{
P.~Griffths and J.~Harris,
``Principles of Algebraic Geometry,''
John Wiley \& Sons Inc., New York 1994.
}


\Title{
\vbox{
\hbox{\tt hep-th/0612095}\vskip -.15cm
\hbox{\tt CERN-PH-TH/2006-256}
}}
{\vbox{
\ifx\answ\bigans
\vskip -5cm
\else
\vskip -4.5cm
\fi
\centerline{\hbox{D-brane monodromies from}}
\vskip 0.3cm
\centerline{\hbox{a matrix-factorization perspective}}
}}
\bigskip
\centerline{Hans Jockers}
 
\bigskip
\centerline{{\it Department of Physics, Theory Division}}
\centerline{{\it CERN, Geneva, Switzerland}}
\bigskip

\ifx\answ\bigans
\vskip -1.0cm
\else
\vskip -.2cm
\fi

\vskip 2.5cm 
\centerline{\bf Abstract}

The aim of this work is to analyze K\"ahler moduli space monodromies of string compactifications. This is achieved by investigating the monodromy action upon D-brane probes, which we model in the Landau-Ginzburg phase in terms of matrix factorizations. The two-dimensional cubic torus and the quintic Calabi-Yau hypersurface serve as our two prime examples.
\bigskip 



\vskip .3in
\Date{\sl {December, 2006}}
\vfill\eject

\newsec{Introduction}
The discovery and investigation of D-branes have given as some insight into the non-perturbative structure of string theory and have improved our understanding of string dualities. However, despite of this success our view upon many aspects of D-branes is still rather limited.

For instance many properties of D-branes in string compactifications are only qualified in certain regions of the string moduli space, such as the geometric regime, where the compactification space is taken to be large compared to the string scale and hence string corrections are suppressed. These scenarios allow us to treat D-branes semi-classically and to apply geometric methods. However, in other regions of the moduli space we cannot neglect stringy quantum corrections \refs{\BrunnerJQ,\DouglasVM,\DiaconescuVP}, and therefore it is necessary to describe D-branes with the machinery of boundary conformal field theory. In principal boundary conformal field theories constitute a suitable description for generic values of the moduli. However, in practice these methods are only applicable at special points in the moduli space, where due to enhanced symmetries the conformal field theory becomes rational and hence solvable \refs{\GepnerVZ,\GepnerQI,\RecknagelSB}. Thus studying D-branes in string compactifications for generic moduli remains a challenge.

Recently matrix factorizations have emerged as yet another tool to study D-branes \refs{\GovindarajanKR,\KapustinBI,\BrunnerDC,\LazaroiuZI,\AshokZB,\HoriBX,\HoriZD} . They model branes in Landau-Ginzburg theories, which describe string compactifications on hypersurfaces in a non-geometric regime of the K\"ahler moduli space \WittenYC. In the context of Landau-Ginzburg models we are still able to continuously vary both bulk complex structure moduli, realized in terms of deformations of the Landau-Ginzburg superpotential, and D-brane moduli, encoded in the matrix factorization \refs{\HoriJA,\BrunnerMT,\GovindarajanIM}. Furthermore, we can even study obstructed moduli and their associated effective superpotentials \refs{\HoriJA,\AshokXQ,\BrunnerMT,\BrunnerTC,\KnappRD,\GovindarajanUY}.

These Landau-Ginzburg theories are believed to flow to an infrared conformal fixed point. Since this flow is rather complicated we use here the framework of topological Landau-Ginzburg theories, which compute quantities invariant with respect to the renormalization group.

The goal of this work is to transport brane probes in the K\"ahler moduli space so as to explore its global structure. But instead of considering an arbitrary path in the moduli space \HerbstWorkP\ (\cf also refs.~\refs{\BrunnerJQ,\DiaconescuVP,\DouglasQW,\GovindarajanVI,\MayrAS,\AspinwallIB}), we are less ambitious and analyze branes as we move along a closed path with base point at the Landau-Ginzburg phase in the K\"ahler moduli space. This corresponds to determining upon matrix factorizations the action of monodromies induced from moduli space singularities. A similar analysis has been carried out in refs.~\refs{\AspinwallPU,\AspinwallDZ,\DistlerYM,\AspinwallJR}, where the large radius point is chosen as a base point and where the monodromies act upon complexes of coherent sheaves.\foot{On the level of D-brane charges monodromies have also been studied in refs.~\refs{\BrunnerJQ,\MayrAS,\BrunnerSK,\BrunnerEG}.} This work should be seen complementary to the large radius results as some of the calculations are more tractable in the language of matrix factorizations.

The outline of the paper is as follows. In section~2 we mainly review matrix factorizations in Landau-Ginzburg orbifolds in order to set our conventions and to introduce the notation. In particular we focus on equivariant matrix factorizations \refs{\AshokZB,\WalcherTX,\GovindarajanIM} and their gradings \WalcherTX, as these properties play an important role in the D-brane monodromy analysis.

Then we turn to the structure of the K\"ahler moduli space of Calabi-Yau hypersurfaces from a gauged linear $\sigma$-model point of view \WittenYC. Typically one obtains three kinds of singularities in the K\"ahler moduli space, namely the large radius, the Landau-Ginzburg and the conifold singularity. In section~3 we investigate in detail the monodromies of these singularities acting upon matrix factorizations. 

In section~4 we employ the developed techniques and study D-brane monodromies on the moduli space of the cubic torus. The matrix factorizations of the cubic torus are well-understood \refs{\BrunnerMT,\GovindarajanIM}, and hence the torus serves as good first example to study the effect of monodromies on matrix factorizations. We also demonstrate that the results are compatible with the expected transformation behavior of D-brane charges. Finally we show the connection of the K\"ahler moduli space as seen from the gauged linear $\sigma$-model \WittenYC\ to the Teichm\"uller space of the two-dimensional torus \LercheCS. 

We turn towards our second example, the quintic Calabi-Yau hypersurface, in section~5. We explicitly address the action of the monodromies upon two types of matrix factorizations of the quintic. Again we verify our results by comparing with the monodromy transformations of the D-brane charges presented in ref.~\refs{\BrunnerJQ,\BrunnerEG}.

In section~6 we present our conclusions and in appendix~A we have collected the open-string cohomology elements used in section~4.

\newsec{D-branes in Landau-Ginzburg orbifold theories}
In order to set the stage for the forthcoming analysis we review the notion of B-type branes in the context of topological Landau-Ginzburg orbifolds. By now it is well-known \refs{\KapustinBI,\BrunnerDC,\AshokZB,\HoriBX,\HoriZD} that B-branes in Landau-Ginzburg theories are given by matrix factorizations of the Landau-Ginzburg superpotential, $W$. In this section we recapitulate the aspects which are important for this work.

\subsec{Matrix factorizations and open-string states}
A B-type brane, $P$, in the topological Landau-Ginzburg theory with homogenous Landau-Ginzburg superpotential, $W(x)$, is realized as matrix, $Q_P$, and a linear involution, $\sigma_P$, \ie $\sigma_P^2={\bf 1}$, such that \refs{\KapustinBI,\BrunnerDC,\AshokZB,\HoriBX,\HoriZD}
\eqn\MFQfact{ Q_P^2(x) \, = \, W(x) \cdot {\bf 1}_{2n\times 2n} \ , \quad 
\sigma_P\,Q_P+Q_P\,\sigma_P \, = \, 0  .}
Here the $2n\times 2n$ matrix, $Q_P$, has polynomial entries in the bulk chiral Landau-Ginzburg fields, $x_\ell$. Furthermore, two matrix factorizations, $(Q_P,\sigma_P)$ and $(Q_{P'},\sigma_{P'})$, are gauge-equivalent, \ie they describe the same brane, if they are related by an invertible $2n\times 2n$ matrix, $U(x)$,\foot{Invertible as a matrix in the ring of polynomials in $x_\ell$.}
\eqn\MFQGauge{ Q_{P'}(x)\,=\,U(x)\,Q_{P}(x)\,U^{-1}(x) \ , \quad 
\sigma_{P'}\,=\,U(x)\,\sigma_P\,U^{-1}(x) \ . }
From a given matrix factorization, $(Q_P,\sigma_P)$, of a brane, $P$, we can immediately construct the matrix factorization, $(Q_{\bar P},\sigma_{\bar P})$, of the anti-brane, $\bar P$, by acting with the operator, $T$:
\eqn\MFBaBQ{T: \ P \mapsto\bar P \ , \quad (Q_P,\sigma_P) \mapsto (Q_P,-\sigma_P) \ . }
Thus the operator, $T$, generates the matrix factorization of the anti-brane.

The physical string states in the topological Landau-Ginzburg theory arise as non-trivial cohomology elements of the BRST operator. For open-string states, $\Theta_{(P,R)}$, of strings stretching from the brane, $P$, to the brane, $R$, the BRST operator is  given by
\eqn\BRSTOp{D_{(P,R)}\Theta_{(P,R)} \,=\, Q_R \Theta_{(P,R)} - \sigma_R\Theta_{(P,R)}\sigma_P\,Q_P \ .}
It is straight forward to check that the BRST operator, $D_{(P,R)}$, squares to zero. 

Furthermore, we observe that the open-string states, $\Theta_{(P,R)}$, split into bosonic states, $\Phi_{(P,R)}$, and fermionic states, $\Psi_{(P,R)}$, which differ by their eigenvalues $\pm 1$ with respect to the involutions of the attached branes:
\eqn\CohBosFermSplit{\sigma_R\Phi_{(P,R)}\sigma_P \,=\,+\,\Phi_{(P,R)} \ , \quad
\sigma_R\Psi_{(P,R)}\sigma_P \,=\,-\,\Psi_{(P,R)} \ . }

In the paper we also use an equivalent description for the matrix factorization, $(Q_P,\sigma_P)$,  which arises as follows: Due to the fact that the matrix, $Q_P$, anti-commutes with the involution, $\sigma_P$, we can always find a gauge in which the involution, $\sigma_P$, takes the block diagonal form $\sigma_P=\Diag{\left({\bf 1}_{n\times n},-{\bf 1}_{n\times n}\right)}$. In this gauge the matrix, $Q_P$, decomposes into two $n\times n$ matrices according to\foot{Note that the block-diagonal form of the involution, $\sigma_P$, corresponds to a partial gauge fixing, which is preserved by gauge transformations~\MFQGauge\ with block-diagonal matrices, $U=\Diag(V_{n\times n},W_{n\times n})$. Here the $n\times n$ matrices, $V_{n\times n}$ and $W_{n\times n}$, are invertible again in the ring of polynomials in $x_\ell$.}
\eqn\MFQBlocks{Q_P(x) \, = \, \pmatrix{0 & J_P(x) \cr E_P(x) & 0 } \ . }
Thus we can alternatively describe the brane, $P$, in terms of the matrix pair, $(J_P, E_P)$, which then fulfills
\eqn\MFfactJE{ J_P(x)\,E_P(x)\,=\,E_P(x)\,J_P(x)\,=\,W(x) \cdot {\bf 1}_{n\times n} \ . }
In this description the operator, $T$, which maps branes to their anti-branes, becomes
\eqn\MFBaB{T:\  P \mapsto \bar P \ , \quad  (J_P,E_P) \mapsto (J_{\bar P},E_{\bar P})=(-E_P,-J_P) \ . }
Moreover, bosonic and fermionic open-string states, $\Phi_{(P,R)}=(\phi_0,\phi_1)$ and $\Psi_{(P,R)}=(\psi_0,\psi_1)$, decompose also into two matrices, and the open-string BRST operator, $D_{(P,R)}$, reads
\eqn\BRSTOpJE{\eqalign{
D_{(P,R)}\Phi_{(P,R)}\,&=\,D_{(P,R)}(\phi_0,\phi_1)\,=\,\left(J_R\phi_0-\phi_1J_P,E_R\phi_1-\phi_0E_P\right) \ , \cr
D_{(P,R)}\Psi_{(P,R)}\,&=\,D_{(P,R)}(\psi_0,\psi_1)\,=\,\left(E_R\psi_0+\psi_1J_P,J_R\psi_1+\psi_0E_P\right) \ . }}

\subsec{R-charge assignments}
For the Landau-Ginzburg model to flow to a non-trivial conformal IR fixed point, it is necessary for the theory to have a (non-anomalous) $U(1)$ R-symmetry. With respect to this $U(1)$ symmetry the bulk Landau-Ginzburg superpotential has R-charge assignment $+2$. Hence for a homogenous superpotential, $W(x)$, of degree $d$ the bulk chiral fields, $x_\ell$, have R-charge $+{2\over d}$.\foot{In this paper we consider only homogenous Landau-Ginzburg superpotentials. The generalization to quasi-homogenous superpotentials is straight forward.}

For Landau-Ginzburg theories with branes it is also necessary to extend the $U(1)$ R-symmetry of the bulk to the boundary. This corresponds to requiring that we can find a $U(1)$ representation, $\rho_P(\theta)$, such that the matrix, $Q_P$, which according to eq.~\MFQfact\ has R-charge $+1$, transforms with respect to the $U(1)$ R-symmetry as \WalcherTX\foot{We always choose a gauge for the matrix factorization, $Q_P$, such that the representation, $\rho(\theta)$, is diagonal and $x$-independent (\cf ref.~\WalcherTX).}
\eqn\RchargeQ{\rho_P(\theta)Q_P(e^{2i{\theta\over d}} x)\rho^{-1}_P(\theta) \,=\,e^{i\theta}Q_P(x) \ .}
Here the representation, $\rho_P(0)$, obeys $\rho_P(0)={\bf 1}_{2n\times 2n}$ and $\rho_P(\pi d)={\bf 1}_{2n\times 2n}$ for even $d$ whereas $\rho_P(2 \pi d)={\bf 1}_{2n\times 2n}$ for odd $d$.

For us it is important that the representations, $\rho_P(\theta)$ and $\rho_R(\theta)$, of the branes, $P$ and $R$, assign also the R-charge, $q_{\Theta_{(P,R)}}$, to the open-string states, $\Theta_{(P,R)}$, 
\eqn\RchargeTheta{\rho_R(\theta)\Theta_{(P,R)}(e^{2i{\theta\over d}} x)\rho^{-1}_P(\theta)=e^{i\theta q_{\Theta_{(P,R)}}}\Theta_{(P,R)}(x) \ .}

\subsec{Equivariant matrix factorizations}
Ultimately we want to study monodromies in the K\"ahler moduli space of Calabi-Yau compactifications. For the compactifications considered in this work the Landau-Ginzburg phase is realized as a Landau-Ginzburg orbifold \WittenYC. The orbifold group, $\IZ_d$, acts on the bulk chiral fields, $x_\ell$, as
\eqn\Orbaction{x_\ell \mapsto \omega^k x_\ell \ , \quad \omega=e^{2\pi i\over d} \ , \quad k\in\IZ_d \ . }
In this context branes are characterized by $\IZ_d$-equivariant matrix factorizations. This means we need to add to the data of the brane, $P$, a $\IZ_d$ representation, $R^P$, such that the matrix, $Q_P$, fulfills the equivariance condition \refs{\AshokZB,\WalcherTX,\GovindarajanIM}:
\eqn\OrbMFRep{R^P(k)Q_P(\omega^k x)R^P(-k)\,=\,Q_P(x) \ . }
In terms of the matrices, $(J_P,E_P)$, the representations, $R^P$, splits into two parts, $R_0^P$ and $R_1^P$, and the equivariance condition~\OrbMFRep\ becomes
\eqn\MFequivJP{\eqalign{
   J_P(x)\,&=\,R_0^P(k)\,J_P(\omega^k x)\,R_1^P(-k) \ , \cr
   E_P(x)\,&=\,R_1^P(k)\,E_P(\omega^k x)\,R_0^P(-k) \ . }}

The expression~\OrbMFRep\ resembles closely the transformation behavior~\RchargeQ\ of the matrix, $Q_P$, with respect to the $U(1)$ R-symmetry. Indeed for irreducible matrix factorizations the representation, $R^P$, are related to the $U(1)$ representation, $\rho_P$, by \WalcherTX
\eqn\OrbRepRcharge{R(k)\,=\,e^{i\pi k \lambda_P} \rho(\pi k) \sigma_P^k \ , \quad a ={\lambda_P d\over 2} \in \IZ \ .  }
Here $\lambda_P$ denotes the grade of the equivariant matrix factorization, which is constraint by $R_P(d)={\bf 1}_{2n\times 2n}$. Thus for each irreducible matrix, $Q_P$, there are $d$ inequivalent $\IZ_d$ representations, $R^{P_a}$, which give rise to $d$ different equivariant branes, $P_a$, in the orbifold theory. Given an equivariant brane, $P$, we simply obtain the other branes, $P_a$, in the same equivariant orbit by
\eqn\MFequiOrbit{R^{P_a}(k)\,=\,\omega^{a k} R^P(k) \ . }

As the representations, $R^P$, distinguishes among the branes in the equivariant orbit we must also adjust the notion of open-string states. Therefore induced from eq.~\OrbMFRep\ we impose on open-string states, $\Theta_{(P,R)}$, the condition
\eqn\OrbMFRep{R^R(k)\Theta_{(P,R)}(\omega^kx)R^P(-k)\,=\,\Theta_{(P,R)}(x) \ . }

\subsec{Gradings of branes} \subseclab\MFGradings
Finally let us discuss one additional refinement in the description of branes. We have seen that branes are equipped with a grade, $\lambda_P$, which, so far, has been ambiguous up to shifts of even integers. As explained in refs.~\refs{\DouglasGI,\DouglasFJ} this ambiguity is not important as long as we analyze the physics of a single brane but becomes relevant for the analysis of open strings stretching between different branes. Thus in order to keep track of this ambiguity, we assign to each brane an integer, $n$, and denote the graded brane by $P[n]$. The grading, $n$, is the integer offset of the grade, $\lambda_P$. If we perform the shift, $\lambda_P\rightarrow\lambda_P-1$, we observe in eq.~\OrbRepRcharge\ that this amounts to changing the sign of the involution, $\sigma_P$, \ie $\sigma_P\rightarrow-\sigma_P$. Thus according to eq.~\MFBaBQ\ the brane, $P[1]$, is the anti-brane of $P[0]$, and hence we identify the operator, $T$, which maps branes to anti-branes, with the translation operator for the integer grading, $n$:
\eqn\MFBaB{T:\  P[n] \mapsto P[n+1] \ . }
Note that in the following we abbreviate the branes, $P[0]$ and $P[1]$, by the short-hand notation, $P$ and $\bar P$.

As a consequence of the interplay of the integer grading, $n$, and the grade, $\lambda_P$, we also obtain the relation
\eqn\MFequivshift{ P_{a+d}[n]\,=\,P_a[n-2] \ . }
Furthermore, for even degrees, $d$, we find that branes and anti-branes are in the same equivariant orbit because the anti-brane, $\bar P_a$, coincides with the brane, $P_{a-d/2}$.

With these definitions at hand we can now assign integer gradings to open-string states. Namely, the grading, $p$, of an open-string state, $\Theta_{(P,R)}$, with R charge, $q_{\Theta_{(P,R)}}$, arises as \WalcherTX
\eqn\OpenGrade{p \, = \, \lambda_R - \lambda_P + q_{\Theta_{(P,R)}} \ . }
For odd and even integers, $p$, the open-string states are bosonic and fermionic respectively. Thus, the integer grading, $p$, is compatible with the statistics of open-string states. We denote the space of open-string states at grading, $p$, by $\Ext^p(P,R)$ and for $p=0$ by $\Hom(P,R)=\Ext^0(P,R)$. Due to eq.~\OpenGrade\ the open-string states at different gradings are related by
\eqn\ExtRel{\Ext^p(P,R)\simeq\Hom(P[-p],R)\simeq\Hom(P,R[p]) \ . }

All those described ingredients are captured in a graded category \refs{\OrlovYP,\OrlovMAG,\AspinwallIB,\LazaroiuWM,\HerbstWorkP}, where the objects are matrix factorizations, the morphisms between objects are open-string states, and finally the shift functor is the operator, $T$. For us it is important to note that in the category of matrix factorizations of topological B-banes, in addition to the gauge equivalences~\MFQGauge, two matrix factorizations are also equivalent if they only differ by blocks of trivial matrix factorizations \refs{\OrlovYP,\HoriJA,\OrlovMAG}
\eqn\MFtrivial{Q_W\,=\,\pmatrix{0 & 1 \cr W & 0} \ , \quad Q_{\bar W}\,=\,\pmatrix{0 & W\cr 1 & 0} \ .}
Physically the trivial matrix factorization, $Q_W$, corresponds to a trivial brane-anti-brane pair, which annihilates to the vacuum.

\newsec{D-brane monodromies in the K\"ahler moduli space}\seclab{\Dmono}
In this section we introduce the tools needed to study D-brane monodromies in the K\"ahler moduli space of hypersurfaces embedded in (weighted) projective spaces. These geometries have a Landau-Ginzburg orbifold phase \refs{\WittenYC,\HerbstWorkP}, in which matrix factorizations describe D-branes, and hence they are suitable to study D-brane monodromies from a matrix factorization perspective.

\subsec{The K\"ahler moduli space and D-brane monodromies}\subseclab{\Kmodspace}
In this paper the cubic torus in $\IC\IP^2$ and the quintic hypersurface in $\IC\IP^4$ serve as our working examples, but the following discussion generalizes to many other Calabi-Yau hypersurfaces as well.

Compactifications of both geometries depend on a single (complexified) K\"ahler modulus and the K\"ahler moduli space becomes singular at three distinct points. There is the large radius point, where the volume of the compactification space becomes infinite, then there is the conifold point, where the (quantum) volume of the hypersurface shrinks to zero size while the (quantum) volume of the lower even dimensional cycles stays finite \GreeneTX, and finally there is the Landau-Ginzburg point, where the singularity in the moduli space arises from a global discrete symmetry of the theory. The structure of the K\"ahler moduli space is schematically depicted in  \lfig\KSpace~(a).
\figinsert\KSpace{
{\bf (a)}\ The figure illustrates the complex one dimensional K\"ahler moduli space of a Calabi-Yau hypersurface with the large radius (LR), the Landau-Ginzburg (LG) and the conifold (C) singularity.\ \ 
{\bf (b)}\ Here we show the three non-trivial loops in the K\"ahler moduli space along which we transport brane probes. The base point of these loops is in the vicinity of the Landau-Ginzburg point, where branes are given in terms of matrix factorizations.}{1.5in}{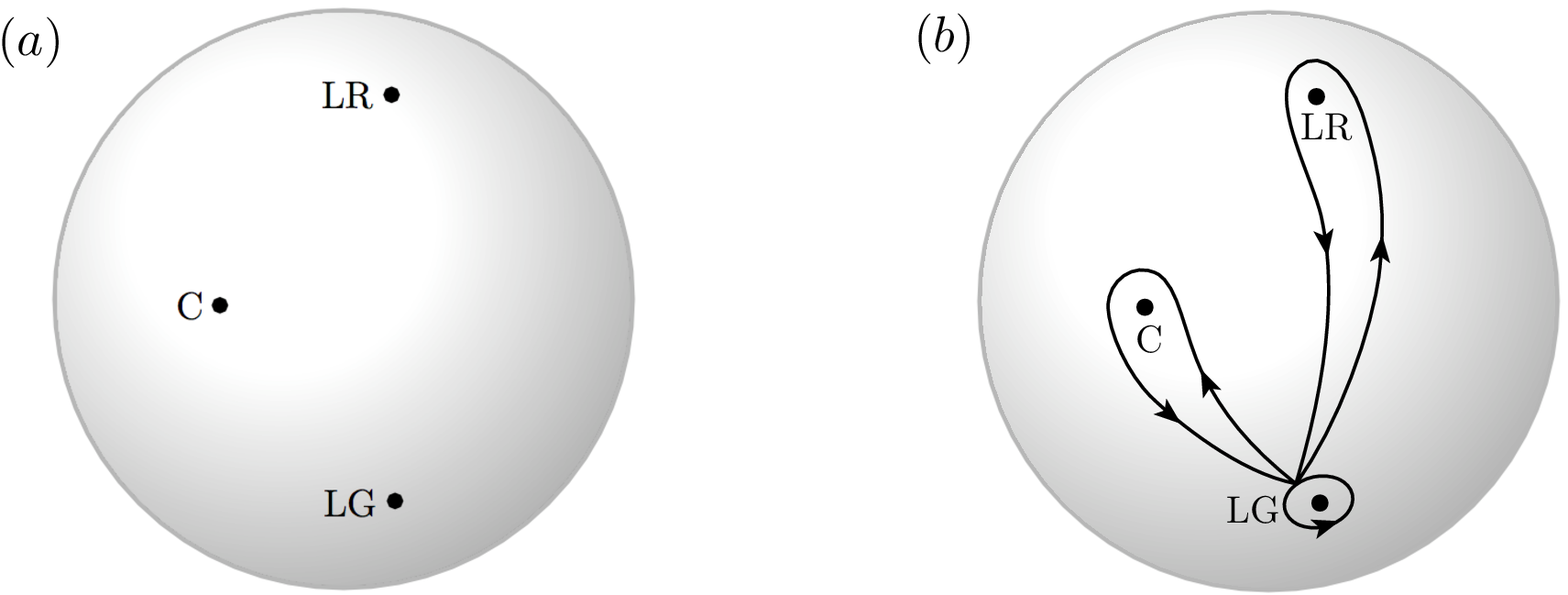}

In the topological B-model the dependence of branes on K\"ahler moduli is rather mild. For instance a brane probe transported along a closed contractible loop is expected to come back unchanged. If, however, the loop is non-contractible, that is to say if we encircle one of the above mentioned singularities, then, in general, the original brane configuration is changed. This, however, does not imply that we get a new theory with different branes. Instead, it just means that the monodromy of the singularity maps individual branes to other branes within the same theory \AspinwallDZ. 

Note that for physical branes there is a stronger dependence on the K\"ahler moduli, as one also has to take into account the notion of $\Pi$-stability \refs{\DouglasAH,\AspinwallPU,\AspinwallDZ}, \ie a physical brane probe can decay as it crosses a line of marginal stability in the K\"ahler moduli space. However, we limit our analysis to topological branes and hence we do not address this issue here. 

Our next task is to discuss the D-brane monodromies arising from the different singularities. As we focus on branes given by matrix factorizations, the base point for the non-contractible loops is located near the Landau-Ginzburg point as depicted in \lfig\KSpace~(b).

\subsec{Landau-Ginzburg point monodromy}
Since we describe branes in the Landau-Ginzburg phase of the $\sigma$-model to the Calabi-Yau hypersurface, the Landau-Ginzburg monodromy is the simplest one in the language of matrix factorizations. At the Landau-Ginzburg point in the K\"ahler moduli space the theory has an enhanced discrete symmetry, which is the orbifold group in the Landau-Ginzburg phase \refs{\LercheCS,\CandelasRM}. Thus encircling the Landau-Ginzburg singularity in the K\"ahler moduli space corresponds to permuting the branes in the equivariant orbit of the Landau-Ginzburg orbifold \refs{\MayrAS,\BrunnerEG,\WalcherTX}. Therefore the monodromy action on the equivariant brane, $P_a$, simply reads
\eqn\LGaction{M_{\rm LG}(P_a)\,=\,P_{a+1} \ , \quad M_{\rm LG}^{-1}(P_a)\,=\,P_{a-1} \ . }

\subsec{Conifold point monodromy}\subseclab{\Cmono}
Next we want to address the monodromy about the conifold point. At the conifold point of Calabi-Yau hypersurfaces the (quantum) volume of the compactification space shrinks to zero size, while the (quantum) volume of lower-dimensional even cycles remains finite \GreeneTX. As a consequence a brane that wraps the compactification space without any lower-dimensional brane charges is massless at the conifold point \refs{\PolchinskiSM,\GreeneTX}. Such a brane, $X$, potentially binds to the transported brane probe, $P$, as follows \AspinwallJR. The mass of a BPS brane is given by the absolute value of its central charge, $Z$, which depends holomorphically on the K\"ahler moduli. Hence at the conifold singularity the central charge, $Z(X)$, of the brane, $X$, is zero and therefore reads in terms of spherical coordinates, $(r,\theta)$, of the K\"ahler moduli space in the vicinity of the singularity
\eqn\ZchargeX{Z(X)\,=\,r\,e^{\pi i \theta} \ . }
On the other hand we assume that the brane probe, $P$, remains massive at the conifold point, and therefore we further assume that close to the conifold point the central charge, $Z(P)$, is to lowest order constant
\eqn\ZchargeP{Z(P)\,=\,c\,e^{\pi i \lambda_ P} \ , }
with some real constant, $c$, and some constant grade, $\lambda_P$. 

The difference of the grades, $\lambda_P-\lambda_X=\lambda_P-\theta$, measures the mass of fermionic open-string states, $\Psi_{(X,P)}$, from brane, $X$, to brane, $P$ \refs{\DouglasAH,\AspinwallPU,\AspinwallDZ}, \ie $\Psi_{(X,P)}$ is massive for $\lambda_P>\lambda_X$ and tachyonic for $\lambda_P<\lambda_X$. As the brane probe, $P$, encircles the conifold point the mass of the open-string state, $\Psi_{(X,P)}$, changes gradually form massive to tachyonic. Thus along the path the pair of branes, $P$ and $X$, becomes unstable and an energetically favored bound state is formed via tachyon condensation. The matrix factorization, $Q_{\rm Con}$, of the condensate with the operator, $\Psi_{(X,P)}$, is easily realized as \refs{\HerbstZM,\GovindarajanIM}
\eqn\ConePX{Q_{\rm Con}\,=\,
\pmatrix{Q_P & \Psi_{(X,P)} \cr 0 & Q_X } \ . }
Here $\Psi_{(X,P)}$ denotes the matrix representative~\CohBosFermSplit\ with respect to the BRST operator~\BRSTOp, and the condensate of the branes, $P$ and $X$,  corresponds to the cone construction, $\Cone\left(\Psi_{(X,P)}: X[-1]\rightarrow P\right)$, with the fermionic operator, $\Psi_{(X,P)}$, as an element of the open-string cohomology group, $\Ext^1(X,P)$.

So far we have skipped an important detail. The grades, $\lambda$, of the central charges, $Z$, correspond in the Landau-Ginzburg phase to the grades of the matrix factorizations discussed in section~\MFGradings. Therefore we have the same integer ambiguity in defining the grade, $\lambda$, from its central charge, $Z$, and the different choices give rise to the integer grading of the brane \refs{\DouglasAH,\AspinwallDZ}:
\eqn\GradeBranes{\lambda_{P[n]}\,=\,\lambda_P-n \ . }
Obviously the integer grading is relevant in the discussion of massive vs. tachyonic open-string operators. The open-string states, which becomes tachyonic along the path around the conifold monodromy, are cohomology elements of $\Ext^1(X,P)$. However, also the other cohomology elements, $\Theta_{(X,P)}$, of $\Ext^p(X,P)$ trigger a condensation process because by eq.~\ExtRel\ they are dual to elements in $\Ext^1(X[1-p],P)$. Hence they generate bound states with the brane, $X[1-p]$, which is also massless at the conifold point.

Thus the brane, $M_{\rm C}(P)$, transformed with respect to the conifold monodromy, arises from condensates of the probe brane, $P$, with the massless branes, $X[n]$. Each cohomology element in $\Ext^1(X[n],P)$, or equivalently each cohomology element in $\Ext^p(X,P)$, gives rise to a tachyonic open-string state along the path around the conifold monodromy and triggers a condensation. The presented heuristic arguments motivate the formula for the conifold monodromy as proposed by Kontsevich \refs{\KontsevichICM,\SeidelIA} 
\eqn\Ccone{M_{\rm C}(P)=\Cone({\rm ev}:\,\hom(X,P)\otimes X\rightarrow P) \ . }
Here $X$ is the brane, which becomes massless at the conifold point, $\hom(X,P)$ denotes the graded complex 
\eqn\homcompl{0\rightarrow \Hom(X,P)\rightarrow \Ext^1(X,P)\rightarrow \Ext^2(X,P) \rightarrow \cdots \ , }
and ${\rm ev}$ is the evaluation map with respect to the elements of $\Ext^p(X,P)$. 

The formula \Ccone\ looks rather superficial. However, in the language of matrix factorizations one can evaluate this equation by a straight forward algorithm:

\noindent
{\it (i)\ \ }Determine the brane, $X$, or rather the matrix factorization, $Q_X$, which becomes massless at the conifold point. This is the D-brane, which in the geometric regime fills the entire compactification space and has no lower-dimensional brane charges \refs{\PolchinskiSM,\GreeneTX}, \eg the pure D$6$-brane for the quintic threefold or the pure D$2$-brane for the two-dimensional cubic torus. 

\noindent
{\it (ii)\ \ }Compute a basis of the open-string cohomology elements, $\Ext^1(X[1-p],P)\simeq\Ext^p(X,P)$. We denote the basis elements by $\Theta^p_{i_p}$, $i_p=1,\ldots,b_p$, $p=0,\ldots,D$, where $b_p$ is the dimension of the cohomology group, $\Ext^p(X,P)$, and $D$ is the complex dimension of the compactification space in the large radius regime. Recall that due to the Calabi-Yau condition we have the relation, $D=d-2$, with the degree, $d$, of the Landau-Ginzburg superpotential.

\noindent
{\it (iii) \ \ }In the last step we construct the cone~\Ccone\ with the matrix representation of the basis, $\Theta^p_{i_p}$, and obtain the matrix factorization of the brane, $M_{\rm C}(P)$:
\eqn\MFcone{Q_{M_{\rm C}(P)}\,=\,\pmatrix{ 
Q_P & \Theta_1^0 & \Theta_2^0 & \cdots & \Theta_{b_D-1}^D & \Theta_{b_D}^D \cr
0 & Q_{X[1]} & 0 & \cdots & 0 & 0 \cr
0 & 0 & Q_{X[1]} & \cdots & 0 & 0 \cr
\vdots & \vdots & \vdots & \ddots & \vdots & \vdots \cr
0 & 0 & 0 & \cdots & Q_{X[1-D]}  & 0 \cr
0 & 0 & 0 & \cdots & 0 & Q_{X[1-D]} } \ . }

Finally let us briefly comment on the inverse conifold monodromy. If we encircle the conifold monodromy with the opposite orientation, then instead of the cohomology elements, $\Theta_{(X,P)}$, their Serre dual cohomology elements, $\hat\Theta_{(P,X)}$, become tachyonic and trigger a bound state formation. Thus the inverse conifold monodromy reads
\eqn\CconeInv{M_{\rm C}^{-1}(P)=\Cone({\rm ev}:\,P\rightarrow\hom(P,X)\otimes X ) \ , }
which translates into the matrix factorization expression
\eqn\MFconeInv{Q_{M_{\rm C}^{-1}(P)}\,=\,\pmatrix{
Q_{X[-1]} & 0 & \cdots & 0 & \hat\Theta_1^0 \cr
0 & Q_{X[-1]} & \cdots & 0 & \hat\Theta_2^0 \cr
\vdots & \vdots & \ddots & \vdots & \vdots \cr
0 & 0 & \cdots & Q_{X[D-1]} & \hat\Theta_{\hat b_D}^D \cr
0 & 0 & \cdots & 0 & Q_P  } \ . }
Here the cohomology elements, $\hat\Theta^q_{i_q}$, $i_q=1,\ldots,\hat b_q$, $q=0,\ldots,D$, constitute a basis of the open-string states, $\Ext^1(P_a,X[q-1])\simeq\Ext^q(P_a,X)$. Due to Serre duality, \ie $\Ext^q(P_a,X)\simeq\Ext^{D-q}(X,P_a)$, the multiplicities, $\hat b_q$ and $b_{D-q}$, coincide, and the cohomology elements, $\hat\Theta^q_{i_q}$, can be chosen to be Serre dual to the elements, $\Theta^{D-q}_{i_{D-q}}$.

One can check that the two monodromy action $\MFcone$ and $\MFconeInv$ are indeed inverse to each other.

\subsec{Large radius point monodromy}
Next we turn to the large radius monodromy, which we deduce indirectly. Encircling first the conifold point and then the Landau-Ginzburg point is equivalent to going around the large radius monodromy in the reverse orientation (\cf \lfig\KSpace~(b)). Therefore from the knowledge of the Landau-Ginzburg and the conifold monodromy we readily compute the large radius monodromy
\eqn\MFLR{
M^{-1}_{\rm LR}(P_a)\,=\,(M_{\rm LG}\circ M_{\rm C})(P_a) \ , \quad
M_{\rm LR}(P_a)\,=\,(M^{-1}_{\rm C}\circ M^{-1}_{\rm LG})(P_a) \ . }
Note that a similar strategy has been employed in refs.~\refs{\AspinwallDZ,\DistlerYM,\AspinwallJR}, where the Landau-Ginzburg monodromy is calculated from the large radius and the conifold monodromy.

\newsec{D-brane monodromies of the cubic torus}
As our first example to study  D-brane monodromies serves the cubic two-dimensional torus, which in the geometric large radius regime arises as the cubic hypersurface
\eqn\Wtorus{W(x)\,=\,x_1^3+x_2^3+x_3^3 - 3\,a\,x_1 x_2 x_3 \ , }
in the projective space, $\IC\IP^2$. Here the parameter, $a$, is the algebraic complex structure modulus, which is related to the flat modulus, $\tau$, of the two-dimensional torus in terms of the modular invariant $j$-function as \LercheCS
\eqn\aModulus{{3 a (a^3+8)\over a^3-1}\,=\,j(\tau) \ . }
In the Landau-Ginzburg phase the relation~\Wtorus\ becomes the cubic superpotential of the Landau-Ginzburg orbifold \WittenYC, where the orbifold group, $\IZ_3$, acts according to eq.~\Orbaction\ as
\eqn\TorusOrbaction{
x_\ell \mapsto \omega^k x_\ell \ , \quad \omega=e^{2\pi i\over 3} \ , \quad k\in\IZ_3 \ . }

The aim of this section is to analyze D-brane monodromies acting upon the `long' and `short' branes, which are represented by matrix factorizations in the Landau-Ginzburg phase of the cubic torus. As we will see the result carries the signature of the underlying gauged linear $\sigma$-model, and we will exhibit the relationship of the monodromies in the linear $\sigma$-model K\"ahler moduli space as depicted in  \lfig\KSpace~(a) to the monodromies in the Teichm\"uller space of the two-dimensional torus.

\subsec{Matrix factorizations of the cubic torus}
The matrix factorizations of the cubic torus are discussed in detail in refs.~\refs{\BrunnerMT,\GovindarajanIM}. Here we briefly review the matrix factorizations of the `long' and the `short' branes, as we will study their monodromy transformations.

The matrix factorization of the three `long'  branes, $L_a$, of the cubic torus is described in terms of the $3\times 3$-matrix pair \BrunnerMT
\eqn\MFLtorus{
J_L\,=\,\pmatrix{ 
{1 \over \alpha_1}G^1_{23} & {1 \over \alpha_3}G^3_{12} & {1 \over \alpha_2}G^2_{13}  \cr
{1 \over \alpha_2}G^{213}_{312} & {1 \over \alpha_1}G^{123}_{213} & {1 \over \alpha_3}G^{312}_{123}  \cr
{1 \over \alpha_3}G^{312}_{213} & {1 \over \alpha_2}G^{213}_{123} & {1 \over  \alpha_1}G^{123}_{312} } \ , \quad
E_L\,=\,\pmatrix{
\alpha_1 \, x_1 & \alpha_2 \, x_3 & \alpha_3 \, x_2   \cr 
\alpha_3 \, x_3 & \alpha_1 \, x_2 & \alpha_2 \, x_1  \cr
\alpha_2 \, x_2 & \alpha_3 \, x_1 & \alpha_1 \, x_3} \ , }
with the quadratic polynomials 
\eqn\DefGPoly{
G^{lmn}_{ijk}\,=\,x_i^2-{\alpha_l^2\over\alpha_m\alpha_n}x_jx_k \ , \quad
G_{jk}^i\,=\,G^{ijk}_{ijk}\,=\,x_i^2-{\alpha_i^2\over\alpha_j\alpha_k}\,x_jx_k \ . }
The parameters, $\alpha_\ell$, are subject to the constraint
\eqn\AlphaConst{0\,=\,\alpha_1^3+\alpha_2^3+\alpha_3^3-3\,a\,\alpha_1\alpha_2\alpha_3 \ , }
and they encode the open-string modulus of the `long' branes, which (projectively) parametrize a continuous family of gauge-inequivalent matrix factorizations. The $U(1)$ representation~\RchargeQ\ of the R-symmetry for the `long' branes reads
\eqn\MFLrho{\rho_L(\theta)\,=\,
\Diag({\bf 1}_{3\times3},\,e^{i\theta\over 3}\,{\bf 1}_{3\times3}) \ . }
and we immediately obtain with eq.~\OrbRepRcharge\ the three equivariant representations
\eqn\MFLequiv{
R^{L_a}_0\,=\,\omega^{ak}\,{\bf 1}_{3\times 3}  \ , \quad
R^{L_a}_1\,=\,\omega^{(a+2)k}\,{\bf 1}_{3\times 3} \ , }
with $\omega\equiv e^{2\pi i\over 3}$. The label, $a$, distinguishes the three `long' branes, $L_a$, in the equivariant orbit of the matrix factorization~\MFLtorus.

Similarly, the `short' branes, $S_a$, of the cubic torus are given by the $2\times 2$-matrix factorization \BrunnerMT
\eqn\MFStorus{
J_S\,=\, \pmatrix{\hphantom{-}L_1 & F_2\cr -L_2 & F_1} \ , \quad
E_S\,=\, \pmatrix{ F_1 & -F_2 \cr L_2 & \hphantom{-}L_1} \ ,}
with the linear entries\foot{We introduce also the linear polynomial, $L_3$, for later convenience.}
\eqn\Mtwolin{L_1\, =\, \alpha_3 x_1 - \alpha_1 x_3 \ , \quad
                   L_2 \, =\, \alpha_3 x_2 - \alpha_2 x_3 \ , \quad
                   L_3 \, = \,\alpha_2 x_1 - \alpha_1 x_2 \ . }
and the quadratic polynomials
\eqn\Mtwoquad{\eqalign{
   F_1 \, &=\, {1\over\alpha_3}\,x_1^2+{\alpha_1\over\alpha_3^2}\,x_1x_3
        -{\alpha_2^2\over\alpha_1\alpha_3^2}\,x_2x_3
        -{\alpha_3\over 2\,\alpha_1\alpha_2}\,x_2x_3-{1\over 2\,\alpha_1}\,x_3^2 \ , \cr
   F_2 \, &=\, {1\over\alpha_3}\,x_2^2+{\alpha_2\over\alpha_3^2}\,x_2x_3
        -{\alpha_1^2\over\alpha_2\alpha_3^2}\,x_1x_3
        -{\alpha_3\over 2\,\alpha_1\alpha_2}\,x_1x_3-{1\over 2\,\alpha_2}\,x_3^2
         \ . }}
Note that, as for the `long' branes, the open-string parameters, $\alpha_\ell$, are constrained by eq.~\AlphaConst, and they also projectively parametrize a continuous family of $2\times 2$ factorizations~\MFStorus. For the `short' branes the $U(1)$ R-symmetry representation becomes
\eqn\MFSrho{\rho_L(\theta)\,=\,
\Diag({\bf 1}_{2\times2},\,e^{-{i\theta\over 3}},e^{i\theta\over 3}) \ , }
and we obtain with eq.~\OrbRepRcharge\ the three equivariant `short' branes, $S_a$, distinguished by their $\IZ_3$~representations
\eqn\MFSequiv{
R^{S_a}_0\,=\,\omega^{ak}\,{\bf 1}_{2\times2} \ , \quad
R^{S_a}_1\,=\,\omega^{ak}\,\Diag(\omega^k,\omega^{2k}) \ ,}
with $\omega\equiv e^{2\pi i\over 3}$.

Finally we introduce the exceptional $4\times 4$-matrix factorization, which contains the pure D$2$-brane in its equivariant orbit \GovindarajanIM
\eqn\MFcantorus{\eqalign{
J_X\,&=\,\pmatrix{
0 & -x_1 & -x_2 & -x_3 \cr
x_1 & 0 & -x_3^2+a\,x_1 x_2 & \phantom{-}x_2^2-a\,x_1 x_3 \cr
x_2 & \phantom{-}x_3^2-a\,x_1 x_2 & 0 & -x_1^2+a\,x_2 x_3 \cr
x_3 & -x_2^2+a\,x_1 x_3 & \phantom{-}x_1^2-a\,x_2 x_3 & 0 } \ , \cr
E_X\,&=\,\pmatrix{
0 & x_1^2-a\,x_2x_3 & x_2^2-a\,x_1 x_3 & x_3^2-a\,x_1 x_2 \cr
-x_1^2+a\,x_2 x_3 & 0 & \phantom{-}x_3 & -x_2 \cr
-x_2^2+a\,x_1 x_3 & -x_3 & 0 & \phantom{-}x_1 \cr
-x_3^2+a\,x_1 x_2 & \phantom{-}x_2 & -x_1 & 0 } \ . }}
This matrix factorization does not depend on any open-string moduli, but it arises in the limit where the
 $3\times 3$ factorization~\MFLtorus\ becomes singular as one of the open-string parameters, $\alpha_\ell$, approaches zero \GovindarajanIM. The $U(1)$ R-symmetry representation \RchargeQ\ is given by
\eqn\MFXrho{\rho_L(\theta)\,=\,
\Diag(e^{2i\theta \over 3},\,{\bf 1}_{3\times3},\,e^{-{i\theta\over 3}},\,e^{i \theta\over 3}\,{\bf 1}_{3\times 3}) \ , }
and the resulting three equivariant representations read
\eqn\MFcanequiv{
R^{X_a}_0\,=\,\omega^{ak}\,\Diag(\omega^k,\,{\bf 1}_{3\times3}) \ , \quad
R^{X_a}_1\,=\,\omega^{ak}\,\Diag(\omega^k,\,\omega^{2k}\,{\bf 1}_{3\times3}) \ , }
which label the branes, $X_a$, in their equivariant orbit.

\subsec{Conifold monodromies of the `long' and `short' branes}\subseclab{\TorusConMono}
Next we turn to the computation of the D-brane monodromies in the language of matrix factorizations. As discussed in section~\Dmono, the monodromy about the Landau-Ginzburg point arises canonically in the context of equivariant matrix factorizations whereas the monodromy about the large radius point is computed indirectly with eq.~\MFLR\ from the Landau-Ginzburg and the conifold monodromy. Therefore we first analyze the monodromy about the conifold point.

Following our recipe for the conifold monodromy outlined in section~\Cmono\ we need to determine the open-string states stretching between the transported brane and the branes, $X[n]$, which become massless at the conifold point.  On the cubic torus we expect the pure D2-brane to become massless.\foot{Strictly speaking the massless brane at the conifold point depends on the path in the K\"ahler moduli space, on which we approach the conifold point. Here we approach the conifold point directly without encircling any other singular points.} In terms of matrix factorizations the D2-brane is realized as one of the branes in the equivariant orbit of the exceptional matrix factorization~\MFcantorus: 
\eqn\MFQXtorus{Q_X\,\equiv\,Q_{X_1} \ .}
The open-string states between the brane, $X$, which becomes massless at the conifold point, and the `long' and `short' branes are depicted in the Quiver diagram \lfig\LSQuiver~\GovindarajanIM. The explicit matrix expressions for these open-string states are collected in Appendix~A. 
\figinsert\LSQuiver{
The quiver diagram displays the fermionic (solid red lines) and bosonic (dashed blue lines) open-string states stretched between the D2-brane, $X$, and the branes, $L_a$ and $S_a$, on the cubic torus. The states, $\Omega_X$, and ${\bf 1}_X$, drawn in light colors, only appear in the open-string moduli space of the `long' brane, $L_1$, where the brane, $L_1$,  is equal to the exceptional D2-brane, $X$, \cf ref.~\GovindarajanIM.}{2.2in}{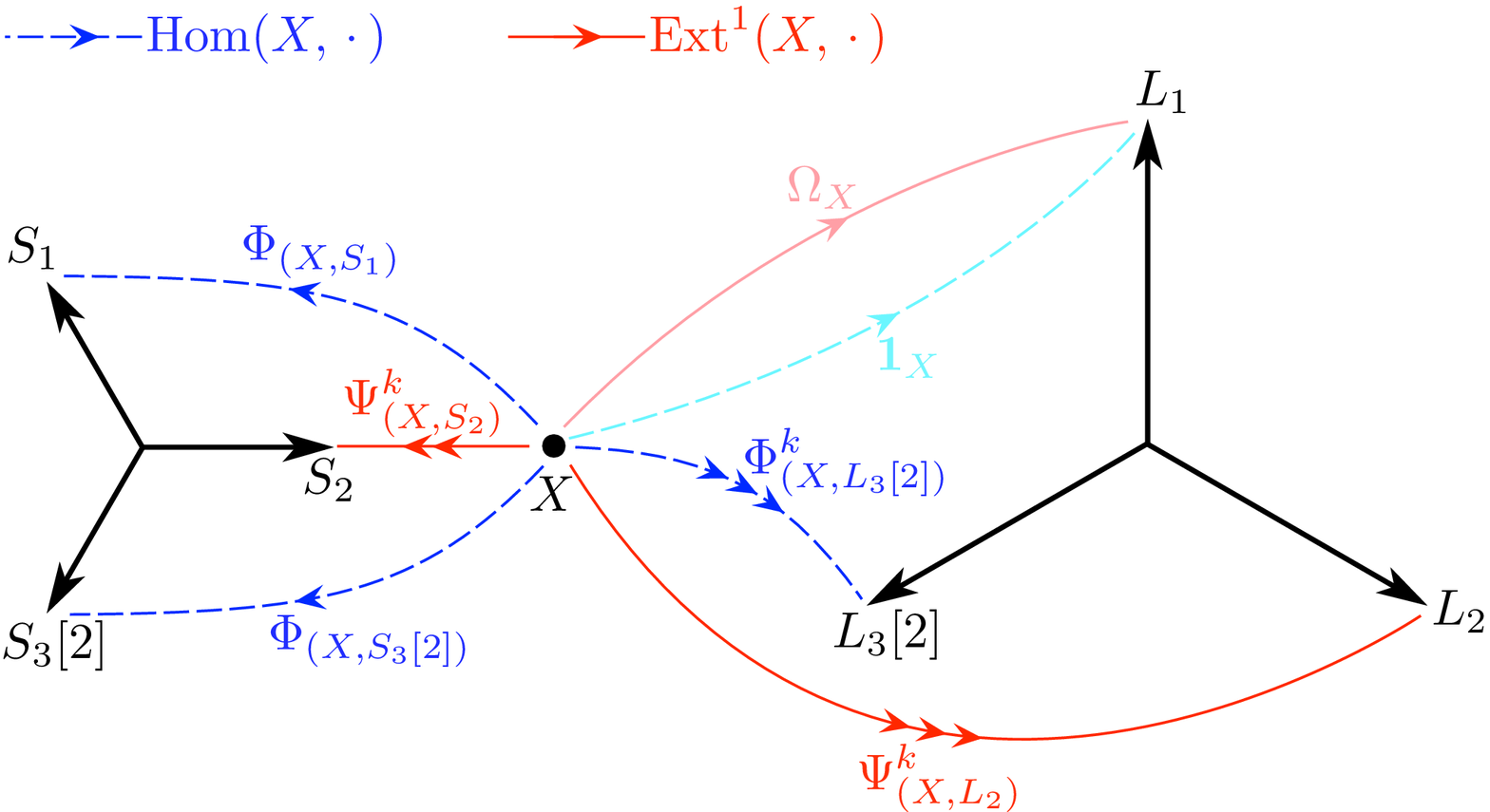}
First we compute the conifold monodromies of the `short' branes. The quiver diagram shows that between the the D2-brane, $X$, and the `short' brane, $S_1$, there is a single bosonic open-string state, $\Phi_{(X,S_1)}$, explicitly given by the matrices~(A.10). Thus applying formula~\MFcone\ for the conifold monodromy we obtain the factorization, $Q_{M_C(S_1)}$, for the transformed brane, $M_C(S_1)$,
\eqn\MonoSone{
Q_{M_C(S_1)}\, = \,\pmatrix{
Q_{S_1} & \Phi_{(X,S_1)} \cr
0 & Q_{\bar X} } \ . }
Here we use the relation, $\Hom(X,S_1)\simeq \Ext^1(\bar X,S_1)$, and by slight abuse of notation, we denote both the bosonic and fermionic open-string states of $\Hom(X,S_1)$ and $\Ext^1(\bar X,S_1)$ by $\Phi_{(X,S_1)}$.

We can further simplify the factorization~\MonoSone\ by applying a gauge transformation~\MFQGauge\ and by subtracting trivial brane-anti-brane pairs~\MFtrivial. In order to keep track of the grading and the equivariant label we also need to simultaneously transform the $U(1)$ R-symmetry representation~\RchargeQ\ and the equivariant representation~\OrbRepRcharge. After a few steps of algebra we obtain that the matrix factorization, $Q_{M_C(S_1)}$, is equivalent to the factorization, $Q_{S_3[2]}$. Thus we have the relation:
\eqn\MonoSoneSimple{M_C(S_1) \, = \, S_3[2] \ . }

Next we consider the monodromy of the `short' brane, $S_2$, about the conifold point. There are two fermionic open-string states, $\Psi^k_{(X,S_2)}$, $k=1,2$, given in eqs.~(A.7) to (A.8), which contribute to the factorization, $Q_{M_C(S_2)}$,
\eqn\MonoStwo{
M_C(S_2)\, = \,
\pmatrix{
Q_{S_2} & \Psi^1_{(X,S_2)} & \Psi^2_{(X,S_2)} \cr
0 & Q_X  & 0 \cr
0 & 0 & Q_X \ } \ .} 
Analogously as before this expression is further simplified by gauge transformations~\MFQGauge\ and by subtracting trivial brane-anti-brane pairs~\MFtrivial. This reduces the $10\times 10$-matrix factorization~\MonoStwo\ to a $7\times 7$-matrix factorization, which explicitly reads: 
\eqn\MonoStwoSimple{\eqalign{
J_{M_C(S_2)}&=
\pmatrix{
\ss 0 &\ss 0 &\ss {x_2\over\alpha_1\alpha_2\alpha_3} 
             &\ss {x_1\over\alpha_1\alpha_2\alpha_3} &\ss {x_3\over\alpha_1\alpha_2\alpha_3} &\ss
             0 &\ss -{x_1\over\alpha_1\alpha_2\alpha_3} \cr
\ss 0 &\ss {x_1\over\alpha_1\alpha_2\alpha_3} &\ss 0 &\ss 
             {x_2\over\alpha_1\alpha_2\alpha_3} &\ss 0 &\ss {x_3\over\alpha_1\alpha_2\alpha_3} 
             &\ss 0 \cr
\ss {1\over\alpha_2}\,G_{23}^1 &\ss -{\alpha _1^2\over\alpha_2^2\alpha_3}\,G_{13}^2 &\ss
             -{1\over\alpha_2}\,G_{12}^3 &\ss -{\alpha_2\over\alpha_1\alpha_3}\,G_{23}^1 &\ss
             {1\over\alpha_2}\,G_{13}^2 &\ss {\alpha_1\over\alpha_3^2}\,G_{12}^3 &\ss H_1 \cr
\ss {1\over\alpha_1}\,G_{13}^2 &\ss -{1\over\alpha_1}\,G_{12}^3 &\ss 
             -{\alpha_2^2\over\alpha_1^2\alpha_3}\,G_{23}^1 &\ss 
             -{\alpha_1\over\alpha_2\alpha_3}\,G_{13}^2 &\ss {\alpha_2\over\alpha_3^2}\,G_{12}^3 &\ss
             {1\over\alpha_1}\,G_{23}^1 &\ss H_2 \cr
\ss {\alpha_1\alpha_2\over\alpha_3^3}\,G_{12}^3 &\ss -{\alpha_2^2\over\alpha_1\alpha_3^2}\,G_{23}^1 
              &\ss -{\alpha_1^2\over\alpha_2\alpha_3^2}\,G_{13}^2 &\ss 
              -{1\over\alpha_3}\,G_{12}^3 &\ss {1\over\alpha_3}\,G_{23}^1 &\ss{1\over\alpha_3}\,G_{13}^2 &\ss
              H_3 \cr
\ss -{x_3\over\alpha_1\alpha_3^2}\,L_1 &\ss \ -{x_2\over\alpha_1\alpha_3^2}\,L_1 &\ss 
              -{x_1\over\alpha_1\alpha_3^2}\,L_1 &\ss 0 &\ss 0 &\ss 0 &\ss 
              -{1\over\alpha_1}\,F_2 \cr
\ss -{x_3\over\alpha_2\alpha_3^2}\,L_2 &\ss -{x_2\over\alpha_2\alpha_3^2}\,L_2 &\ss 
              -{x_1\over\alpha_2\alpha_3^2}\,L_2 &\ss 0 &\ss 0 &\ss 0 &\ss 
              {1\over\alpha_2}\,F_1 } \ ,\cr
E_{M_C(S_2)}&=
\pmatrix{
\ss \alpha _2^3\,G_{23}^1 &\ss  \alpha _1^3\,G_{13}^2 &\ss \alpha_2\,x_1  &\ss x_2\,\alpha_1 &\ss 0 &\ss 
        {\alpha_3^2\over 2\alpha_2}\,M_2 &\ss {\alpha_3^2\over 2\alpha_1}\,M_1 \cr
\ss {\alpha_1^2 \alpha_2^2\over\alpha _3}\,G_{12}^3  &\ss \alpha _1 \alpha _2 \alpha _3\,G_{23}^1 &\ss 
        0 &\ss -\alpha_1\,x_3  &\ss 0 &\ss \alpha_2^2\,x_3 &\ss -\alpha_2\,M_2 \cr
\ss \alpha _1 \alpha _2 \alpha _3\,G_{13}^2 &\ss {\alpha _1^2 \alpha _2^2\over \alpha_3}\,G_{12}^3  &\ss
         -\alpha_2\,x_3  &\ss 0 &\ss 0 &\ss -\alpha_1\,M_1 &\ss \alpha_1^2\,x_3  \cr
\ss \alpha_1\alpha_2\alpha_3\,G_{23}^1 &\ss \alpha_1\alpha_2\alpha_3\,G_{13}^2 &\ss 0 &\ss 0 &\ss 
         -\alpha_3\,x_3 &\ss \alpha_1\alpha_2\,x_3  &\ss \alpha_2\alpha_3\,x_1 \cr
\ss \alpha_1\alpha_2\alpha_3\,G_{12}^3 &\ss \alpha_2^3\,G_{23}^1  &\ss 
         \alpha_2\,x_2 &\ss 0 &\ss \alpha_3\,x_1 &\ss \alpha_1^2\,x_2 &\ss -\alpha_1\,M_3 \cr
\ss \alpha_1^3\,G_{13}^2  &\ss \alpha_1 \alpha_2 \alpha_3\,G_{12}^3 &\ss 
         0 &\ss \alpha_1\,x_1 &\ss \alpha_3\,x_2 &\ss -\alpha_2\,M_3 &\ss \alpha_2^2\,x_1 \cr
\ss 0 &\ss 0 &\ss 0 &\ss 0 &\ss 0 &\ss -\alpha_1\,L_2 &\ss \alpha_2\,L_1}\ . }}
Here we write the entries of the matrices in terms of the linear and quadratic terms~\Mtwolin\ and \Mtwoquad\ and the polynomials  
\eqn\DefPoly{\eqalign{
H_1\,&=\,
\left({\alpha_2\over\alpha_3}-{\alpha_3^2\over 2 \alpha_2^2}\right){x_1^2\over\alpha_1}
+{\alpha_1\,x_2^2\over\alpha_2\alpha_3}-{\alpha_1\,x_3^2\over\alpha_2^2}
+{x_1x_2\over\alpha_3}-{\alpha_3\,x_1x_3\over\alpha_2^2}
-{\alpha_1\alpha_3\,x_2x_3\over 2\alpha_2^3} \ , \cr
H_2\,&=\,
-{\alpha_2\,x_1^2\over\alpha_1\alpha_3}+{\alpha_3^2\,x_2^2\over 2\alpha_1^2\alpha_2}
+{\alpha_2\,x_3^2\over\alpha_1^2}-{x_1x_2\over\alpha_3}
-\left({\alpha_2\over\alpha_3^2}-{\alpha_2\alpha_3\over2\alpha_1^3}\right) x_1x_3
+{\alpha_3\,x_2x_3\over\alpha_1^2} \ , \cr
H_3\,&=\,
{\alpha_1\,x_1^2\over\alpha_3^2}-{\alpha_2\,x_2^2\over\alpha _3^2}+{x_3^2\over2\alpha_3}
-\left({\alpha_3^2\over 2}+{\alpha_2^3\over\alpha_3}
-{\alpha_1^3\over\alpha_3}\right){x_1x_2\over\alpha_1\alpha_2\alpha_3}
+{\alpha_1^2\,x_1x_3\over\alpha_3^3}-{\alpha_2^2\,x_2x_3\over\alpha_3^3}  \ , \cr
M_1\,&=\,\alpha_3\,x_1+\alpha_1\,x_3 \ , \quad\
M_2\,=\,\alpha_3\,x_2+\alpha_2\,x_3 \ , \quad\
M_3\,=\,\alpha_2\,x_1+\alpha_1\,x_2 \ . }}
The $U(1)$ R-symmetry representation and the equivariant representation for the matrix factorization~\MonoStwoSimple\ for the brane, $M_C(S_2)$, becomes
\eqn\MFCStwo{\rho_{M_C(S_2)}(\theta)\,=\,
\Diag({\bf 1}_{2\times 2},\,e^{-{2i\theta\over 3}}\,{\bf 1}_{5\times 5},\,e^{-{i\theta\over3}}\,{\bf 1}_{7\times 7}) \ , }
and
\eqn\MStwoSimpleEQ{
R^{M_C(S_2)}_0 \, = \, \Diag(\omega^{2k}\,{\bf 1}_{2\times 2},\,\omega^k\,{\bf 1}_{5\times 5}) \ , \quad
R^{M_C(S_2)}_1 \, = \, {\bf 1}_{7\times 7} \ . }

For the `short' brane, $S_3[2]$, the quiver diagram~\lfig\LSQuiver\ reveals one bosonic open-string states, $\Phi_{(X,S_3[2])}$. By shifting the grades along the lines of eq.~\ExtRel\ this bosonic open-string state is mapped into the cohomology group, $\Ext^1(\bar X[-2],S_3)$. Then it describes a fermionic open-string state stretching from the anti-D2-brane, $\bar X[-2]$, to the `short' brane, $S_3$. Therefore the conifold monodromy acts upon the `short' brane, $S_3$, as
\eqn\MonoSthree{
Q_{M_C(S_3)}\, = \,\pmatrix{
Q_{S_3} & \Phi_{(X,S_3[2])} \cr
0 & Q_{\bar X[-2]} } \ , }
with the matrix~(A.11) for $\Phi_{(X,S_3[2])}$. We simplify this $6\times 6$-matrix factorization with an appropriate gauge transformation~\MFQGauge, and we subtract one trivial brane-anti-brane pair~\MFtrivial\ to arrive at the $5\times 5$-matrix factorization:
\eqn\MonoSthreeSimple{\eqalign{
J_{M_C(S_3)}&=\pmatrix{
\ss -\alpha_2L_1 &\ss 0 &\ss \alpha_1\left(\alpha_3^2x_3-\alpha_2^2x_2\right) &\ss
     \alpha_1^2\alpha_2x_1-\alpha _3^3x_2  &\ss 
     {\alpha_3^4\over 2\alpha_1}x_1+{\alpha_3^3\over 2}x_3-\alpha_2^2\alpha_3x_2 \cr
\ss \alpha_1L_2 &\ss 0 &\ss \alpha_1\alpha_2^2x_2-\alpha_3^3x_1 &\ss
     \alpha_2\left(\alpha_3^2x_3-\alpha_1^2x_1\right) &\ss 
     {\alpha _3^4\over 2 \alpha_2}x_2+{\alpha _3^3\over 2}x_3-\alpha_1^2\alpha_3x_1 \cr
\ss 0 &\ss {1\over\alpha_1}G_{23}^1 &\ss 0 &\ss -\alpha_3x_3 &\ss \alpha_2x_2 \cr
\ss 0 &\ss {1\over\alpha_2}G_{13}^2 &\ss \alpha_3x_3 &\ss 0 &\ss -\alpha_1x_1 \cr
\ss 0 &\ss {1\over\alpha_3}G_{12}^3 &\ss -\alpha_2x_2 &\ss \alpha_1x_1 &\ss 0 } \ , \cr 
E_{M_C(S_3)}&=\pmatrix{
\ss -{1\over\alpha_2}F_1 &\ss {1\over\alpha_1}F_2 &\ss
      K_1 &\ss K_2 &\ss K_3 \cr
 \ss 0 &\ss 0 &\ss \alpha_1 x_1 &\ss \alpha_2x_2 &\ss \alpha_3x_3 \cr
 \ss -{\alpha_1x_1\over\alpha_2\alpha_3^4}L_2 &\ss -{x_1\over\alpha_3^4}L_1 &\ss
        -{\alpha_1x_1\over\alpha_3^3}L_2 &\ss 
        {1\over\alpha_3}G_{12}^3+{\alpha_1\over\alpha_3^2}G_{13}^1 &\ss
        -{1\over\alpha_2}G_{13}^2-{\alpha_1\alpha_2\over\alpha_3^3}G_{12}^1 \cr
\ss -{x_2\over\alpha_3^4}L_2 &\ss -{\alpha_2x_2\over\alpha_1\alpha_3^4}L_1 &\ss 
        -{1\over\alpha_3}G_{12}^3-{\alpha_2\over\alpha_3^2}G_{23}^2 &\ss 
        {\alpha_2x_2\over\alpha_3^3}L_1 &\ss
        {1\over\alpha_1}G_{23}^1+{\alpha_1\alpha_2\over\alpha_3^3}G_{12}^2 \cr
\ss -{x_3\over\alpha_2\alpha_3^3}L_2 &\ss -{x_3\over\alpha_1\alpha_3^3}L_1 &\ss 
        {1\over\alpha_2}G_{13}^2+{\alpha_2\over\alpha_3^2}G_{23}^3 &\ss 
        -{1\over\alpha_1}G_{23}^1-{\alpha_1\over\alpha_3^2}G_{13}^3 &\ss 
        -{x_3\over\alpha_3^2}L_3 } \ . }}
The entries of this matrix factorization are abbreviated by the polynomials~\Mtwolin, \Mtwoquad\ and \DefGPoly\ and by
\eqn\DefPolytwo{\eqalign{
K_1\,&=\,
-{\alpha_3^3\,x_2^2\over2\alpha_1\alpha_2^2}-{\alpha_3\,x_3^2\over 2\alpha_1}
+{\alpha_1\,x_1 x_2\over\alpha_2}-{\alpha_3^2\,x_1x_3\over 2\alpha_1^2}
+\left({\alpha_2^2\over\alpha_3}-{\alpha_3^2\over2\alpha_2}\right){x_2 x_3\over\alpha_1} \ , \cr
K_2\,&=\,
-{\alpha_3^3\,x_1^2\over 2\alpha_1^2\alpha_2}-{\alpha_3\,x_3^2\over 2\alpha_2}+{\alpha_2\,x_1x_2\over\alpha_1}
+\left({\alpha_1^2\over\alpha_3}-{\alpha_3^2\over 2\alpha_1}\right){x_1x_3\over\alpha_2}
-{\alpha_3^2\,x_2 x_3\over 2\alpha_2^2} \ , \cr
K_3\,&=\,
{\alpha_1\,x_1^2\over\alpha _3}+{\alpha_2\,x_2^2\over \alpha_3}
-{\alpha_3^2\,x_1x_2\over\alpha_1\alpha_2}-{\alpha_3\,x_1x_3\over 2 \alpha_1}-{\alpha_3\,x_2x_3\over 2\alpha_2} \ . }}
The $U(1)$ R-symmetry representation of the transformed `short' brane, $M_C(S_3)$, which is associated to the simplified factorization~\MonoSthreeSimple, is given by
\eqn\MFCSthree{\rho_{M_C(S_3)}(\theta)\,=\, \Diag({\bf 1}_{5\times 5},
\,e^{-{i\theta\over 3}},\,e^{i\theta\over 3},\,e^{-{i\theta\over 3}}\,{\bf 1}_{3\times 3} )  \ , }
whereas the $\IZ_3$-equivariant representation becomes
\eqn\MSthreeSimpleEQ{
R^{M_C(S_2)}_0 \, = \, {\bf 1}_{5\times 5}  \ , \quad
R^{M_C(S_2)}_1 \, = \, \Diag(\omega^k,\,\omega^{2 k},\,\omega^k\,{\bf 1}_{3\times 3}) \ . }

Now we turn to the analysis of the monodromy about the conifold point acting on the three equivariant `long' branes. The quiver diagram~\lfig\LSQuiver\ shows again the open-string spectrum, which is relevant to evaluate the conifold monodromy for the `long' branes. The matrix representations of these open-string states are collected in Appendix~A.

First we consider the `long' brane, $L_1$. At a generic point in the open-string moduli space there are no open-string states stretching between the pure D2-brane, $X$, and the `long' brane, $L_1$. Therefore the monodromy about the conifold point leaves the `long' brane, $L_1$, simply invariant:
\eqn\MonoLone{M_C(L_1) \, = \, L_1 \ . }
However, if we choose the open-string modulus such that the factorization~\MFLtorus\ of $L_1$ becomes singular, \ie if one of the open-string parameters, $\alpha_\ell$, in the factorization~\MFLtorus\ approaches zero, then, as discussed in ref.~\GovindarajanIM, the factorization of the `long' brane turns into the exceptional matrix factorization~\MFcantorus\ of the brane, $X_1$. Hence at this exceptional point in the open-string moduli space the `long' brane, $L_1$, coincides with the pure D2-brane, $X_1$, and as a consequence the (bosonic) identity operator, ${\bf 1}_X$, and its fermionic Serre dual operator, $\Omega_X$, appear in the open-string spectrum (\cf~\lfig\LSQuiver).\foot{Note that as the bosonic and the fermionic open-string states arise simultaneously, the index of the open-string spectrum remains invariant over the open-string moduli space.} Thus at this point in the open-string moduli space the conifold monodromy acts upon the `long' brane, $L_1\equiv X_1$, as
\eqn\MonoLExceptone{
Q_{M_C(X_1)}\,=\,
\pmatrix{
Q_X & \Omega_X & {\bf 1}_X \cr
0 & Q_X & 0 \cr
0 & 0 & Q_{\bar X}} \ . }  
This factorization actually simplifies to again the factorization, $Q_{X_1}$, by applying a gauge transformations~\MFQGauge, which allows us to drop eight trivial brane-anti-brane pairs~\MFtrivial. Thus at all points in the open-string moduli space the relation~\MonoLone\ holds because also the exceptional `long' brane, $X_1$, undergoes the conifold monodromy unchanged.

The open-string spectrum between the D2-brane, $X$, and the `long' brane, $L_2$, consists for all open-string moduli of three fermionic open-string states, $\Psi^k_{(X,L_2)}$, $k=1,2,3$ (\cf~\lfig\LSQuiver), given in eqs.~(A.1) to (A.3). Therefore with eq.~\MFcone\ we find for the conifold monodromy of the brane, $L_2$, the factorization
\eqn\MonoLtwo{
Q_{M_C(L_2)}\,  = \,
\pmatrix{
Q_{L_2} & \Psi^1_{(X,L_2)} & \Psi^2_{(X,L_2)} & \Psi^3_{(X,L_2)} \cr
0 & Q_X & 0 & 0 \cr
0 & 0 & Q_X & 0 \cr
0 & 0 & 0 & Q_X } \ .
}
Analogously to the previous examples due to gauge transformations~\MFQGauge\ and due to equivalences arising from trivial brane-anti-brane pairs~\MFtrivial\ this $15\times 15$-matrix factorization simplifies to a $9\times 9$-matrix factorization
\eqn\MonoLtwoSimple{\eqalign{
{\ss J_{M_C(L_2)}}&\ss=\pmatrix{
\sss x_3 &\sss x_1 &\sss x_2 &\sss 0 &\sss 0 &\sss 0 &\sss 0 &\sss 0 &\sss 0 \cr
\sss 0 &\sss 0 &\sss 0 &\sss x_2 &\sss x_3 &\sss x_1 &\sss 0 &\sss 0 &\sss 0 \cr
\sss 0 &\sss 0 &\sss 0 &\sss 0 &\sss 0 &\sss 0 &\sss x_1 &\sss x_2 &\sss x_3 \cr
\sss 0 &\sss {G^2_{13}\over\alpha_1} &\sss -{G^1_{23}\over\alpha_1} &\sss {\alpha_3G^2_{13}\over\alpha_1\alpha_2} &\sss 0 
                   &\sss -{G^3_{12}\over\alpha_3} &\sss -{\alpha_3G^1_{23}\over\alpha_1^2} 
                   &\sss {\alpha_2G^3_{12}\over\alpha_1\alpha_3} &\sss 0 \cr
\sss -{G^2_{13}\over\alpha_2} &\sss 0 &\sss {G^3_{12}\over\alpha_2} &\sss -{G^1_{23}\over\alpha_1} 
                   &\sss {\alpha_1G^3_{12}\over\alpha_2\alpha_3} &\sss 0 &\sss 0 &\sss -{\alpha_1G^2_{13}\over\alpha_2^2} 
                   &\sss {\alpha_3G^1_{23}\over\alpha_1\alpha_2}  \cr
\sss {G^1_{23}\over\alpha_3} &\sss -{G^3_{12}\over\alpha_3} &\sss 0 &\sss 0 &\sss -{G^2_{13}\over\alpha_2} 
                    &\sss -{\alpha_3G^1_{23}\over\alpha_1^2} &\sss {\alpha_1G^2_{13}\over\alpha_2\alpha_3} &\sss 0 
                    &\sss -{\alpha_2G^3_{12}\over\alpha_3^2} \cr
\sss {x_1x_2\over\alpha_3} &\sss {\alpha_2x_1x_3\over\alpha_1^2} &\sss {\alpha_3x_1^2\over\alpha_1\alpha_2} 
                    &\sss {\alpha_3x_1x_3\over\alpha_1^2} &\sss {x_1^2\over\alpha_2} &\sss {\alpha_2x_1x_2\over\alpha_1\alpha_3} 
                    &\sss {\alpha_3^2x_1^2\over\alpha_1^2\alpha_2} &\sss -{T_3\over\alpha_1}  &\sss {x_2^2\over\alpha_1} \cr
\sss {\alpha_1x_2^2\over\alpha_2\alpha_3} &\sss {x_2 x_3\over\alpha_1} &\sss {\alpha_3x_1x_2\over\alpha_2^2} 
                    &\sss {\alpha_3x_2x_3\over\alpha_1\alpha_2} &\sss {\alpha_1x_1x_2\over\alpha_2^2}
                    &\sss {x_2^2\over\alpha_3} &\sss {x_3^2\over\alpha_2} &\sss {\alpha_1^2x_2^2\over\alpha_2^2\alpha_3} 
                    &\sss  -{T_1\over\alpha_2} \cr
\sss {\alpha_1x_2x_3\over\alpha_3^2} &\sss {\alpha_2x_3^2\over\alpha_1\alpha_3} &\sss {x_1x_3\over\alpha_2} 
                    &\sss {x_3^2\over\alpha_1} &\sss {\alpha_1x_1x_3\over\alpha_2\alpha_3} &\sss {\alpha_2x_2x_3\over\alpha_3^2} 
                    &\sss -{T_2\over\alpha_3} &\sss {x_1^2\over\alpha_3} &\sss {\alpha_2^2x_3^2\over\alpha_1\alpha_3^2}
} \ , \cr 
{\ss E_{M_C(L_2)}}&\ss=\pmatrix{
\ss G^3_{12} &\ss -{\alpha_2\over\alpha_1}G^1_{23} &\ss -{\alpha_1\over\alpha_2}G^2_{13}  
               &\ss 0  &\ss -\alpha_2x_2  &\ss \alpha_3x_1 &\ss 0 &\ss 0 &\ss 0 \cr
\ss G^1_{23} &\ss {\alpha_3\over\alpha_2}G^2_{13} &\ss -{\alpha_2\over\alpha_3}G^3_{12}   &\ss
               \alpha_1x_2 &\ss 0 &\ss -\alpha_3x_3 &\ss 0 &\ss 0 &\ss 0 \cr
\ss G^2_{13} &\ss -{\alpha_1\over\alpha_3}G^3_{12}  &\ss -{\alpha_3\over\alpha_1}G^1_{23} &\ss
               -\alpha_1x_1 &\ss \alpha_2x_3 &\ss 0 &\ss 0 &\ss 0 &\ss 0 \cr
\ss 0 &\ss G^2_{13} &\ss 0 &\ss 0 &\ss -\alpha_1x_1 &\ss \alpha_2x_3 &\ss 0 &\ss -\alpha_3x_1 &\ss \alpha_1x_3 \cr
\ss 0 &\ss G^3_{12} &\ss 0 &\ss \alpha_3x_1 &\ss 0 &\ss -\alpha_2x_2 &\ss \alpha_2x_1 &\ss 0 &\ss -\alpha_1x_2 \cr
\ss 0 &\ss G^1_{23} &\ss 0 &\ss -\alpha_3x_3 &\ss \alpha_1x_2 &\ss 0 &\ss -\alpha_2x_3 &\ss  \alpha_3x_2 &\ss 0 \cr
\ss 0 &\ss 0 &\ss G^1_{23} &\ss 0 &\ss 0 &\ss 0 &\ss 0 &\ss \alpha_2x_3 &\ss -\alpha_3x_2 \cr
\ss 0 &\ss 0 &\ss G^2_{13} &\ss 0 &\ss 0 &\ss 0 &\ss -\alpha_1x_3 &\ss 0 &\ss  \alpha_3x_1 \cr
\ss 0 &\ss 0 &\ss G^3_{12} &\ss 0 &\ss 0 &\ss 0 &\ss  \alpha_1x_2 &\ss -\alpha_2x_1 &\ss 0
} \ . }}
In these matrices we introduce in addition to the polynomials~\DefGPoly\ the quadratics
\eqn\DefPolythree{
T_1\,=\,x_1^2-{\alpha_1^3+\alpha_2^3\over\alpha_1\alpha_2\alpha_3}\,x_2x_3 \ , \quad
T_2\,=\,x_2^2-{\alpha_2^3+\alpha_3^3\over\alpha_1\alpha_2\alpha_3}\,x_1x_3 \ , \quad 
T_3\,=\,x_3^2-{\alpha_1^3+\alpha_3^3\over\alpha_1\alpha_2\alpha_3}\,x_1x_2 \ . }
The $U(1)$ R-symmetry representation for the matrix factorization~\MonoLtwoSimple\ becomes
\eqn\MLRohRep{\rho_{M_C(L_2)}\,=\,
\Diag({\bf 1}_{3\times 3},\,e^{-{2i\theta\over 3}}\,{\bf 1}_{6\times 6},\,e^{-{i\theta\over 3}}\,{\bf 1}_{9\times 9}) \ , }
whereas the $\IZ_3$ equivariant representation turns out to be
\eqn\MLtwoSimpleEQ{
R^{M_C(L_2)}_0\,=\, \Diag(\omega^{2k}\,{\bf 1}_{3\times 3},\,\omega^k\,{\bf 1}_{6\times 6}) \ , \quad
R^{M_C(L_2)}_1\,=\, {\bf 1}_{9\times 9} \ . }

Finally let us turn to the conifold monodromy acting on the remaining `long' brane, $L_3$. From the Quiver diagram~\lfig\LSQuiver\ we extract that there are three bosonic open-string states, $\Phi^k_{(X,L_3[2])}$, $k=1,2,3$, given by eqs.~(A.4) to (A.6). Similarly to the analysis of the `short' brane, $S_3$, using eq.~\ExtRel\ we map these bosonic open-string states to fermionic open-string states stretching between the anti-D2-brane, $\bar X[-2]$, and the `long' brane, $L_3$. Then, with slight abuse of notation for these fermionic states, we write the conifold monodromy action upon the brane, $L_3$, as
\eqn\MonoLthree{
Q_{M_C(L_3)}\,  = \,
\pmatrix{
Q_{L_3} & \Phi^1_{(X,L_3[2])} & \Phi^2_{(X,L_3[2])} & \Phi^3_{(X,L_3[2])} \cr
0 & Q_{\bar X[-2]} & 0 & 0 \cr
0 & 0 & Q_{\bar X[-2]} & 0 \cr
0 & 0 & 0 & Q_{\bar X[-2]} } \ .}
With the help of gauge transformations and factorization equivalences this $15\times 15$-matrix factorization reduces to a $12\times 12$-matrix factorization, which in terms of the quadratic polynomials~\DefGPoly\ and
\eqn\DefGTildePoly{
\tilde G^1_{23}\,=\,G^1_{23}-{\alpha_2^2\over\alpha_1\alpha_3}x_2x_3 \ , \quad
\tilde G^2_{13}\,=\,G^2_{13}-{\alpha_3^2\over\alpha_1\alpha_2}x_1x_2 \ , \quad
\tilde G^3_{12}\,=\,G^3_{12}-{\alpha_1^2\over\alpha_2\alpha_3}x_1x_2 \ , }
can be written as:
\eqn\MonoLthreeSimple{\eqalign{
&J_{M_C(L_3)}=\pmatrix{
\ssk 0 &\sskk 0 &\sskk 0 &\sskk -{\alpha_2x_3\over\alpha_1\alpha_3} &\sskk {x_2\over\alpha_2} &\sskk 0 &\sskk 0 
           &\sskk {\alpha_3x_1\over\alpha_1\alpha_2} &\sskk {x_3\over\alpha_3} &\sskk {x_1\over\alpha_1} &\sskk 0 
           &\sskk -{\alpha_1x_2\over\alpha_2 \alpha_3} \skcr
\ssk 0 &\sskk 0 &\sskk 0 &\sskk {x_1\over\alpha_3} &\sskk 0 &\sskk -{\alpha_3x_2\over\alpha_1\alpha_2} 
           &\sskk -{\alpha_1x_3\over\alpha_2\alpha_3} &\sskk {x_2\over\alpha_1} &\sskk 0 &\sskk 0 
           &\sskk -{\alpha_2x_1\over\alpha_1\alpha_3} &\sskk {x_3\over\alpha_2} \skcr
\ssk 0 &\sskk 0 &\sskk 0 &\sskk -{\alpha_1x_2\over\alpha_2\alpha_3} &\sskk 0 &\sskk {x_3\over\alpha_1} &\sskk {x_1\over\alpha_2} 
           &\sskk -{\alpha_2x_3\over\alpha_1\alpha_3} &\sskk 0 &\sskk 0 &\sskk {x_2\over\alpha_3} 
           &\sskk -{\alpha_3x_1\over\alpha _1\alpha_2} \skcr
\ssk G^1_{23} &\sskk 0 &\sskk -{\alpha_2^2G_{12}^3\over\alpha_3^2} &\sskk 0 &\sskk -x_3 &\sskk x_2 &\sskk 0 &\sskk 0 &\sskk 0 
           &\sskk 0 &\sskk 0 &\sskk 0 \skcr
\ssk G^2_{13} &\sskk 0 &\sskk -{\alpha_2G_{23}^1\over\alpha_1} &\sskk x_3 &\sskk 0 &\sskk -x_1 &\sskk 0 &\sskk 0 &\sskk 0 
           &\sskk 0 &\sskk 0 &\sskk 0 \skcr
\ssk G^3_{12} &\sskk 0 &\sskk -{\alpha_2G_{13}^2\over\alpha_3} &\sskk -x_2 &\sskk x_1 &\sskk 0 &\sskk 0 &\sskk 0 &\sskk 0 
           &\sskk 0 &\sskk 0 &\sskk 0 \skcr
\ssk -{\alpha_2G_{12}^3\over\alpha_1} &\sskk G^1_{23} &\sskk 0 &\sskk 0 &\sskk 0 &\sskk 0 &\sskk 0 &\sskk -x_3 &\sskk x_2 
           &\sskk 0 &\sskk 0 &\sskk 0 \skcr
\ssk -{\alpha_3^2G_{23}^1\over\alpha_1^2} &\sskk G^2_{13} &\sskk 0 &\sskk 0 &\sskk 0 &\sskk 0 &\sskk x_3 &\sskk 0 &\sskk -x_1 
           &\sskk 0 &\sskk 0 &\sskk 0 \skcr
\ssk -{\alpha_3G_{13}^2\over\alpha_2} &\sskk G^3_{12} &\sskk 0 &\sskk 0 &\sskk 0 &\sskk 0 &\sskk -x_2 &\sskk x_1 &\sskk 0 
           &\sskk 0 &\sskk 0 &\sskk 0 \skcr
\ssk 0 &\sskk -{\alpha_1G_{12}^3\over\alpha_3} &\sskk G_{23}^1 &\sskk 0 &\sskk 0 &\sskk 0 &\sskk 0 &\sskk 0 &\sskk 0 &\sskk 0 
           &\sskk -x_3 &\sskk x_2 \skcr
\ssk 0 &\sskk -{\alpha_3G_{23}^1\over\alpha_2} &\sskk G_{13}^2 &\sskk 0 &\sskk 0 &\sskk 0 &\sskk 0 &\sskk 0 &\sskk 0 
           &\sskk x_3 &\sskk 0 &\sskk -x_1 \skcr
\ssk 0 &\sskk -{\alpha_1^2G_{13}^2\over\alpha_2^2} &\sskk G^3_{12} &\sskk 0 &\sskk 0 &\sskk 0 &\sskk 0 &\sskk 0 &\sskk 0 
           &\sskk -x_2 &\sskk x_1 &\sskk 0 } \ , \cr 
&E_{M_C(L_3)}=\cr
&\ \pmatrix{
\ssk 0 &\sskk 0 &\sskk 0 &\sskk x_1 &\sskk x_2 &\sskk x_3 &\sskk 0 &\sskk 0 &\sskk 0 &\sskk 0 &\sskk 0 &\sskk 0 \skcr
\ssk 0 &\sskk 0 &\sskk 0 &\sskk 0 &\sskk 0 &\sskk 0 &\sskk x_1 &\sskk x_2 &\sskk x_3 &\sskk 0 &\sskk 0 &\sskk 0 \skcr
\ssk 0 &\sskk 0 &\sskk 0 &\sskk 0 &\sskk 0 &\sskk 0 &\sskk 0 &\sskk 0 &\sskk 0 &\sskk x_1 &\sskk x_2 &\sskk x_3 \skcr
\ssk \alpha_2x_1x_2 &\sskk \alpha_3x_1^2 &\sskk \alpha_1x_1x_3 &\sskk 0 &\sskk G_{12}^3 &\sskk -x_2^2 
           &\sskk -{\alpha_2x_1x_3\over\alpha_3} &\sskk 0 &\sskk 0 &\sskk 0 &\sskk -{\alpha_1x_2^2\over\alpha_3} 
           &\sskk {\alpha_2G^1_{23}\over\alpha_1} \skcr
\ssk \alpha_2x_2^2 &\sskk \alpha_3x_1x_2 &\sskk \alpha_1x_2x_3 &\sskk -G^3_{12} &\sskk 0 
           &\sskk \tilde G_{23}^1 &\sskk -{\alpha_2x_2x_3\over\alpha_3} &\sskk 0 
           &\sskk 0 &\sskk {\alpha_1G_{13}^2\over\alpha_3} &\sskk -{\alpha_2^2x_2x_3\over\alpha_3^2} 
           &\sskk -{\alpha_2^2G^3_{12}\over\alpha_3^2} \skcr
\ssk \alpha_2x_2x_3 &\sskk \alpha_3x_1x_3 &\sskk \alpha_1x_3^2 &\sskk G^2_{13} &\sskk -G^1_{23} 
           &\sskk -{\alpha_2^2x_3^2\over\alpha_1\alpha_3} &\sskk -{\alpha_2x_3^2\over\alpha_3} &\sskk 0 &\sskk 0 
           &\sskk -{\alpha_2G_{23}^1\over\alpha_1} &\sskk -{\alpha_2x_1x_2\over\alpha_1} &\sskk 0 \skcr
\ssk \alpha_3x_1x_3 &\sskk \alpha_1x_1x_2 &\sskk \alpha_2x_1^2 &\sskk 0 &\sskk -{\alpha_3G^2_{13}\over\alpha_2} 
           &\sskk -{\alpha_3x_2x_3\over\alpha_2} &\sskk -{\alpha_3^2x_1^2\over\alpha_1\alpha_2} 
           &\sskk G^3_{12} &\sskk -G^2_{13} &\sskk 0 &\sskk -{\alpha _3x_1^2\over\alpha_1} &\sskk 0 \skcr
\ssk \alpha_3x_2x_3 &\sskk \alpha_1x_2^2 &\sskk \alpha_2x_1x_2 &\sskk -{\alpha_3G^2_{13}\over\alpha_2} 
           &\sskk 0 &\sskk -{\alpha_2x_3^2\over\alpha_1} &\sskk -x_3^2 &\sskk 0 &\sskk G^1_{23} &\sskk 0 
           &\sskk -{\alpha_3x_1x_2\over\alpha_1} &\sskk 0 \skcr
\ssk \alpha_3x_3^2 &\sskk \alpha_1x_2x_3 &\sskk \alpha_2x_1x_3 &\sskk -{\alpha_3^2G^1_{23}\over\alpha_1^2} 
           &\sskk {\alpha_2G^3_{12}\over\alpha_1} &\sskk -{\alpha_3^2x_1x_3\over\alpha_1^2} 
           &\sskk \tilde G_{13}^2 &\sskk -G^1_{23} &\sskk 0 &\sskk 0 &\sskk -{\alpha_3x_1x_3\over\alpha_1} &\sskk 0 \skcr
\ssk \alpha_1x_1^2 &\sskk \alpha_2x_1x_3 &\sskk \alpha_3x_1x_2 &\sskk 0 &\sskk 0 &\sskk -{\alpha_1x_1x_2\over\alpha_2} 
           &\sskk -{\alpha_1^2x_1x_2\over\alpha_2^2} &\sskk -{\alpha_1^2G^2_{13}\over\alpha_2^2} 
           &\sskk {\alpha_3G^1_{23}\over\alpha_2} &\sskk 0 &\sskk \tilde G^3_{12} &\sskk -G_{13}^2 \skcr
\ssk \alpha_1x_1x_2 &\sskk \alpha_2x_2x_3 &\sskk \alpha_3x_2^2 &\sskk 0 &\sskk 0 &\sskk -{\alpha_1x_2^2\over\alpha_2} 
           &\sskk -{\alpha_1x_1x_3\over\alpha_3} &\sskk 0 &\sskk -{\alpha_1G^3_{12}\over\alpha_3} 
           &\sskk -G^3_{12} &\sskk -{\alpha_1^2x_2^2\over\alpha_2\alpha_3} &\sskk G^1_{23} \skcr
\ssk \alpha_1x_1x_3 &\sskk \alpha_2x_3^2 &\sskk \alpha_3x_2x_3 &\sskk 0 &\sskk 0 &\sskk -{\alpha_1x_2x_3\over\alpha_2} 
           &\sskk -{\alpha_3x_1^2\over\alpha_2} &\sskk {\alpha_1G^3_{12}\over\alpha_3} &\sskk 0 &\sskk G^2_{13} 
           &\sskk -x_1^2 &\sskk 0  } \ . }}
Furthermore, the $U(1)$ R-symmetry representation reads
\eqn\MCLthreeRohRep{\rho_{M_C(L_3)}\,=\,\Diag({\bf 1}_{12\times12},\,e^{i\theta\over3}{\bf 1}_{3\times3},\,e^{-{i\theta\over3}}{\bf 1}_{9\times9} ) \ , }
and the $\IZ_3$ equivariant representation is given by
\eqn\MLthreeSimpleEQ{
R^{M_C(L_3)}_0\,=\, {\bf 1}_{12\times12} \ , \quad R^{M_C(L_3)}_1\,=\,\Diag(\omega^{2k}\,{\bf 1}_{3\times3},\,\omega^k\,{\bf 1}_{9\times9})  \ . }

This completes the calculation of the conifold monodromy acting on the `long' and `short' branes. In the next section these results serve as our starting point to analyze the remaining monodromies  and in discussing global properties of the K\"ahler moduli space.

\subsec{D-brane monodromies of the `long' and `short' branes}\subseclab{\TorusMono}
With the analysis of the monodromy about the conifold point performed in the previous section we can now discuss the remaining K\"ahler moduli space monodromies. The monodromy about the Landau-Ginzburg point from the perspective of equivariant matrix factorizations is straight forward as it simply shifts the equivariant label of the brane. In practice this amounts to multiplying the equivariant $\IZ_3$ representation of the factorization with $\omega\equiv e^{2\pi i\over 3}$ along the lines of eq.~\MFequiOrbit. Hence the Landau-Ginzburg monodromy acts upon the defining data of the brane, $P$, simply by
\eqn\TorusLGmono{M_{\rm LG}:\,
\pmatrix{Q_P\cr \rho_P \cr R^P(k)} \mapsto
\pmatrix{Q_{M_{\rm LG}(P)}\cr \rho_{M_{\rm LG}(P)} \cr R^{M_{\rm LG}(P)}(k) }
\,=\,\pmatrix{Q_P\cr \rho_P \cr \omega^k R^P(k)} \ . }
Here $P$ represents any equivariant brane, in particular any of the `long' and `short' branes, $S_a$ and $L_a$.

Along the lines of eq.~\MFLR\ we combine the conifold and the Landau-Ginzburg monodromy to deduce the action of the inverse large radius monodromy.\foot{In order to get the large radius monodromy one needs to compute according to eq.~\MFLR\ the inverse conifold monodromy. In this work we do not present this computation explicitly as it does not lead to further insight compared to the computation of the inverse large radius monodromy.} Thus together with eq.~\TorusLGmono\ we obtain for the inverse large radius monodromy of the brane, $P$,
\eqn\TorusLRmono{M^{-1}_{\rm LR}:\,
\pmatrix{Q_P\cr \rho_P \cr R^P(k)} \mapsto
\pmatrix{Q_{M^{-1}_{\rm LR}(P)}\cr \rho_{M^{-1}_{\rm LR}(P)} \cr R^{M^{-1}_{\rm LR}(P)}(k) }
\,=\,\pmatrix{Q_{M_C(P)}\cr \rho_{M_C(P)} \cr \omega^k R^{M_C{P}}(k)} \ . }
%

\tabinsert\tbRRTorus{Conifold and large radius monodromies acting on the `long' and `short' branes, $L_a$ and $S_a$, of the two-dimensional torus together with their RR charges.}
{\centerline{\vbox{
\offinterlineskip
\tabskip=0pt\halign{
\vrule height12pt depth6pt#\tabskip=2.5pt plus 1fil\strut &\hfil#\hfil&\vrule# &\hfil#\hfil&\vrule# &\hfil#\hfil&\vrule# 
    &\hfil#\hfil&\vrule# &\hfil#\hfil&\vrule# &\hfil#\hfil &\tabskip=0pt\vrule#\cr
\noalign{\hrule}
&Brane $P_a$ && $\ch_{\rm LR}(P_a)$ && $M_{\rm C}(P_a)$ && $\ch_{\rm LR}(M_{\rm C}(P_a))$
     && $M_{\rm LR}^{-1}(P_a)$ && $\ch_{\rm LR}(M_{\rm LR}^{-1}(P_a))$ &\cr
\noalign{\hrule height1pt}
&$L_1$ && $(1,0)$    && $L_1$                  && $(1,0)$   && $L_2$                             && $(1,-3)$ &\cr
&$L_2$ && $(1,-3)$   && $M_{\rm C}(L_2)$ && $(4,-3)$   && $M_{\rm LR}^{-1}(L_2)$    && $(1,-6)$ &\cr
&$L_3$ && $(-2,3)$   && $M_{\rm C}(L_3)$ && $(-5,3)$   && $M_{\rm LR}^{-1}(L_3)$    && $(-2,9)$ &\cr
\noalign{\hrule}
&$S_1$ && $(0,1)$    && $S_3[2]$                  && $(-1,1)$    && $S_1$                         && $(0,1)$   &\cr
&$S_2$ && $(1,-2)$   && $M_{\rm C}(S_2)$  && $(3,-2)$   && $M_{\rm LR}^{-1}(S_2)$ && $(1,-5)$  &\cr
&$S_3$ && $(-1,1)$   && $M_{\rm C}(S_3)$  && $(-2,1)$   && $M_{\rm LR}^{-1}(S_3)$ && $(-1,4)$  &\cr
\noalign{\hrule} }}}}

The Landau-Ginzburg monodromy and the large radius monodromy does not introduce new matrix factorizations, $Q$, but instead modifies the equivariant representation, $R$, of the branes. The transformation behavior of the `long' and `short' branes is summarized in \ltab\tbRRTorus.

In this table we have also included the large radius RR charges. These charges are computed by a set of disk correlators, where we insert a basis of RR ground states in the bulk and where the brane data enters in the boundary condition of the disk. In the context of matrix factorizations these disk correlators are computed by the residue formula \refs{\KapustinGA,\HerbstAX,\WalcherTX}
\eqn\DiskRes{
\braket{l;\alpha}{P} 
\,=\,{1\over r_l!}\Res_{W_l}\left[\phi^\alpha_l 
      \Str\left((R^P)^l(\partial Q_{P,l})^{\wedge r_l}\right)\right] \ . }
Here $\ket{P}$ is the boundary state of the brane, $P$, and $\ket{l;\alpha}$ denotes a basis of RR ground states, which are labeled by the twisted sectors, $l$, whereas the label, $\alpha$, distinguishes further the RR ground states in each twisted sector. The integer, $r_l$, denotes the number of untwisted fields, $x_\ell$, in each twisted sector, $l$. The details of the disk correlator are explained in ref.~\WalcherTX. For us, however, it is important to note that all correlators~\DiskRes\ for $r_l\ne 0$ vanish for both the cubic torus and the quintic Calabi-Yau hypersurface. Hence we only need to evaluate the correlators with $r_l=0$, for which the residue formula reduces to \WalcherTX
\eqn\DiskResZero{\braket{l;0}{P}\,=\,\Str\left[ (R^P)^l \right] \ . }

For the cubic superpotential~\Wtorus\ all untwisted fields vanish, \ie $r_l=0$, in the sectors, $l=1,2$, and hence the only potentially non-vanishing disk correlators on the cubic torus are $\braket{1;0}{P}$ and $\braket{2,0}{P}$. Thus we readily obtain the RR charge vector, $\ch_{\rm LG}(P)$, 
\eqn\RRLGchargeTorus{\ch_{\rm LG}(P) \,=\, \left( \braket{1;0}{P}, \braket{2;0}{P} \right) 
\,=\,\left( \Str\left[R^P\right],\,\Str\left[ (R^P)^2 \right]\right) \ . }
Note that these charges are given in the basis which arises naturally at the Landau-Ginzburg point in the K\"ahler moduli space. However, in order to gain some geometric intuition we want to relate these charges to the large radius charge vector, $\ch_{\rm LR}(P)$,
\eqn\RRLRchargeTorus{\ch_{\rm LR}(P) \,=\, \left(r, c_1 \right) \ . }
Here, $r$ is the D2-brane charge and $c_1$ is the D0-brane charge. Geometrically these two quantities correspond to the rank and the first Chern class of the bundle date associated in the large radius regime to the brane, $P$. The two charge vectors, $\ch_{\rm LG}(P)$ and $\ch_{\rm LR}(P)$, are related by the $2\times 2$-transformation matrix, $\Xi$, 
\eqn\TransCharge{\ch_{\rm LR}(P) \, = \, \ch_{\rm LG}(P)\cdot \Xi  \ . }

Thus in order to calculate the large radius charges of any equivariant factorization, we need first to determine the matrix, $\Xi$. We know that the pure D2-brane in the large radius regime is represented by the brane, $X_1$, and hence has the charge, $\ch_{\rm LR}(X_1)=(1,0)$. Furthermore, the matrix factorization, $X_2$, is in the same equivariant orbit and has according to refs.~\refs{\BrunnerMT,\GovindarajanIM} the large radius charges, $\ch_{\rm LR}(X_2)=(1,-3)$. By comparing with the Landau-Ginzburg charges~\DiskResZero,
\eqn\RRLGonetwo{
\ch_{\rm LG}(X_1)\,=\,\left(3-3\omega^2, 3-3\omega \right) \ , \quad
\ch_{\rm LG}(X_2)\,=\,\left(-3\omega+3\omega^2, 3\omega-3\omega^2 \right) \ , }
we readily determine the transformation matrix, $\Xi$, to be
\eqn\TransXi{\Xi\,=\,\pmatrix{
{1\over 3(1-\omega)} & {\omega^2\over 3} \cr
{1\over 3(1-\omega^2)} & -{1\over 3(\omega+1)} } \ .}
With the explicit expression for the transformation matrix, $\Xi$, we can now compute with eqs.~\RRLGchargeTorus\ and \TransCharge\ the large radius charges for all the factorizations collected in \ltab\tbRRtorus.

Let us know take a closer look at the transformation behavior of the individual branes listed in \ltab\tbRRtorus. In the previous section we have already seen that the `long' brane, $L_1$, is not affected by the conifold monodromy. Therefore the large radius monodromy maps the `long' brane, $L_1$, to the `long' brane, $L_2$. 

The `short' brane, $S_1$, is the pure D0-brane in agreement with its large radius RR~charges, and its open-string modulus parametrizes the position of the D0-brane on the two-dimensional torus \refs{\BrunnerMT,\GovindarajanIM}. With respect to the large radius monodromy the brane, $S_1$, remains invariant. This is precisely the transformation behavior we expect because the large radius monodromy corresponds to an integer shift of the B-field. But on the point-like worldvolume of the brane the B-field has no support and therefore the D0-brane, $S_1$, remains unchanged. Note also the interplay of gradings among the different monodromies. The conifold monodromy shifts the grade of $S_1$ by two to $S_3[2]$ (\cf eq.~\MonoSoneSimple), which is again compensated by yet another shift~\MFequivshift\ of $-2$ resulting from the Landau-Ginzburg monodromy. Hence the inverse large radius monodromy~\MFLR, as arising from the composition of the other two monodromies, does not modify the grading of the D0-brane, $S_1$.

For all the branes listed in \ltab\tbRRtorus\ we observe that the large radius monodromy transforms the large radius RR~charges as 
\eqn\LRtensor{\otimes\cL^{-3}\,:\,(r,c_1) \rightarrow (r, c_1-3r) \ . }
This transformation behavior is natural from the gauged linear $\sigma$-model point of view, in which the large radius monodromy shifts the B-field of the cubic torus by the two form, $\Theta$, induced from the generator of $H^2(\IC\IP^2,\IZ)$ of the ambient space, $\IC\IP^2$. Note, however, that the generator of $H^2(T^2,\IZ)$ is the two-form, ${1\over 3}\Theta$, instead of the induced two-from, $\Theta$. Therefore the large radius monodromy in the linear $\sigma$-model corresponds to tensoring with the line bundle, $\cL^3$, where $\cL$ is the line bundle of the torus with first Chern number one. Hence, encircling the inverse large radius monodromy is associated with the tensor product by the bundle, $\cL^{-3}$, which generates the transformation~\LRtensor\ for the RR~charges.

Although the gauged linear $\sigma$-model favors a shift of the B-field induced from the ambient space, we would expect that the large radius monodromy of the two-dimensional torus is generated by tensoring with the line bundle, $\cL$. However, the moduli space, as analyzed from the gauged linear $\sigma$-model, does not reveal the whole structure of the Teichm\"uller space of the two-dimensional torus. The relationship to the Teichm\"uller moduli space is further analyzed in the next section.

\subsec{Teichm\"uller and gauge linear $\sigma$-model moduli space of the cubic torus}
The K\"ahler moduli space of the two-dimensional torus is parametrized by the fundamental domain of its Teichm\"uller space (\cf \lfig\TSpace~(a)). Due to the identifications in the fundamental domain the Teichm\"uller space has three singularities, namely a $\IZ_4$-orbifold point, $P_4$, a $\IZ_6$-orbifold point, $P_6$, and the point, $P_\infty$, of infinite order \LercheCS.
\figinsert\TSpace{
{\bf (a)}\ The figure shows the fundamental domain of the Teich\-m\"uller moduli space. Its boundaries are identified according to the black arrows. These identifications generate the three singularities, $P_4$, $P_6$, $P_\infty$, indicated in red.\ \
{\bf (b)}\ Here we illustrate the K\"ahler moduli space of the cubic torus as seen from the gauged linear $\sigma$-model, which is a fourfold cover of the fundamental domain. In blue we show the path associated to the large radius~(LR) and the conifold~(C) monodromy in the gauged linear $\sigma$-model.}{2.0in}{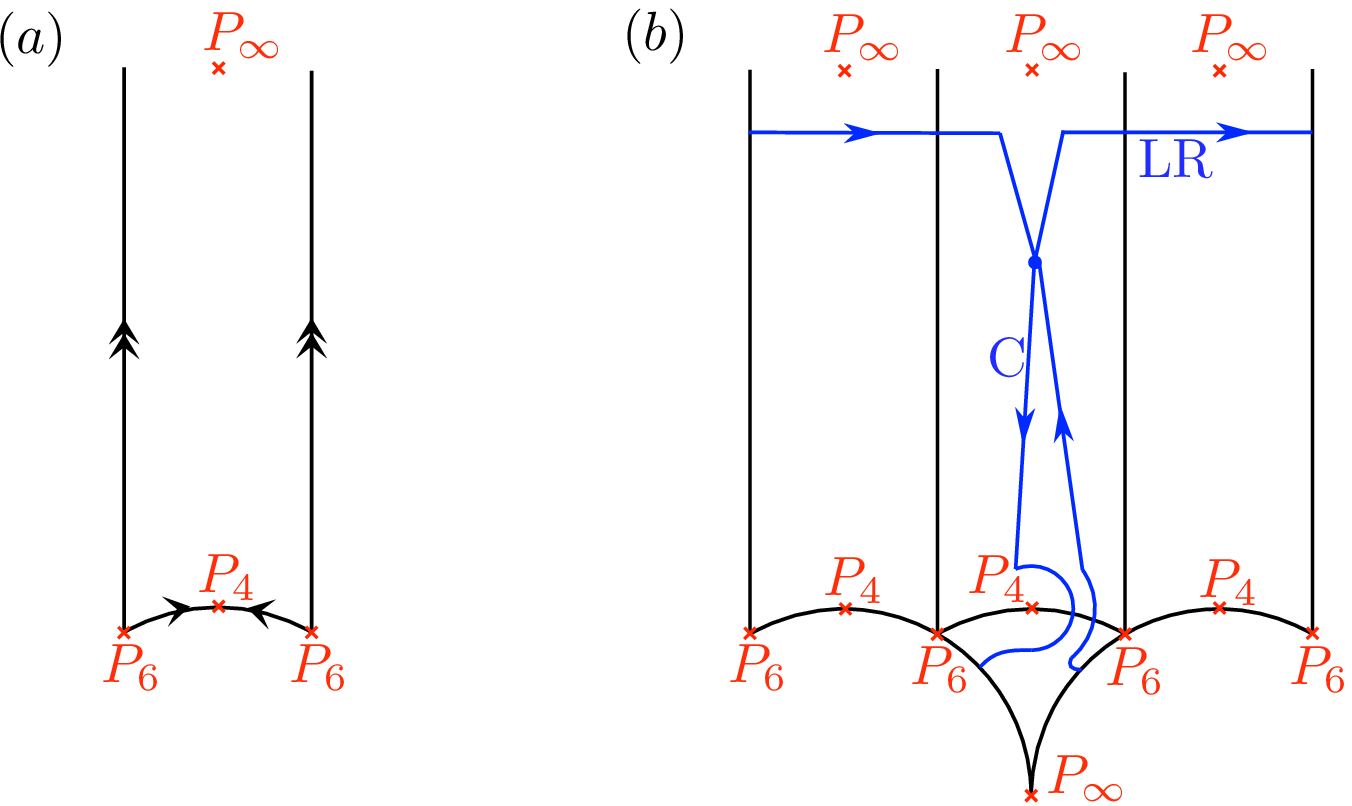}
Here we are interested how these singularities generate monodromies acting upon the RR~charges of the toroidal B-branes. The monodromies, however, are most easily determined on the mirror torus, where the B-branes with RR~charges, $(r,c_1)$, are mapped to A-branes realized as special Lagrangian submanifolds with winding numbers, $(p,q)$ \PolishchukDB. On the mirror side the monodromies are generated by encircling the corresponding singularities in the complex structure moduli space, which, for the torus, is identical to the Teichm\"uller space depicted in \lfig\TSpace~(a). Thus we are able to determine geometrically the effect of the monodromies by simply tracing the fate of the winding numbers as we encircle the singularities in the complex structure moduli space, and we obtain
\eqn\TeichTrans{
P_\infty\,=\,\pmatrix{1 & -1 \cr 0 & \phantom{-}1 } \ , \quad
P_4\,=\,\pmatrix{\phantom{-}0 & 1 \cr -1 & 0 } \ , \quad
P_6\,=\,\pmatrix{0 & -1 \cr 1 & \phantom{-}1 } \ , }
with $P_4^4\equiv{\bf 1}_{2\times 2}$ and $P_6^6\equiv{\bf 1}_{2\times2}$. 

Finally we want to make the connection to the gauged linear $\sigma$-model K\"ahler moduli space. In the previous section we have shown that the large radius monodromy shifts the first Chern number by multiples of three. Hence, so as to generate the large radius monodromy of the cubic torus we should encircle three times the singularity, $P_\infty$, in the Teichm\"uller space. Furthermore, taking again a look at \ltab\tbRRTorus\ we observe that the conifold monodromy shifts the rank, $r\rightarrow r-c_1$, by the first Chern class, $c_1$. Thus we should also identify the conifold monodromy with the singularity, $P_\infty$. However, compared to the large radius monodromy the roles of the rank, $r$, and the Chern number, $c_1$, are interchanged, and hence we identify the conifold point with the singularity, $P_\infty$, which in the covering space of the Teichm\"uller space is S-dual to the large radius singularity, $P_\infty$. To summarize we can view the K\"ahler moduli space of the gauged linear $\sigma$-model of the cubic torus as the fourfold cover of the Teichm\"uller space depicted in \lfig\TSpace~(b), where three fundamental domains are related by translations and where one fundamental domain is S-dual to one of the three others.  

Let us now qualitatively relate the Teichm\"uller monodromies to the linear $\sigma$-model monodromies. In \lfig~\TSpace~(b) the paths around the large radius and the conifold monodromy are also drawn and they gives rise to the relations $M_{\rm LR}=P_\infty^3$ and $M_{\rm C}=P_4P_\infty P_4^{-1}$. The Landau-Ginzburg monodromy from the linear $\sigma$-model point of view must then be comprised of the monodromies around the singularities which are traversed if we deform in \lfig\TSpace~(b) the conifold contour into the large radius contour. This procedure yields $M_{\rm LG}=(P_4P_6)^3P_4^2P_6P_4^{-1}$. Using the matrices~\TeichTrans\ we explicitly obtain:
\eqn\SigmaTrans{
M_{\rm LR}\,=\,\pmatrix{1 & -3 \cr 0 & \phantom{-}1} \ , \quad
M_{\rm C}\,=\,\pmatrix{1 & 0 \cr 1 & 1 } \ , \quad
M_{\rm LG}\,=\,\pmatrix{-2 & 3 \cr -1 & 1 } \ . }
It is easy to check that $M_{\rm LG}^3\equiv{\bf 1}_{2\times2}$ and that the matrices reproduce the RR~charge transformations listed \ltab\tbRRTorus. 

\newsec{D-brane monodromies of the quintic Calabi-Yau hypersurface}
The quintic Calabi-Yau threefold serves as our second example in studying D-brane monodromies. At the large radius point the quintic hypersurface is realized as the zero locus of the quintic polynomial,
\eqn\Wquintic{W(x)\,=\sum_{i=1}^5 x_i^5 - 5\,\psi\,x_1x_2x_3x_4x_5 \ , }
in the complex four-dimensional projective space, $\IC\IP^4$. The $101$ complex structure deformations of the quintic threefold are captured by homogeneous deformations of the polynomial~\Wquintic. For simplicity we exhibit here only the dependence on a single complex structure modulus expressed in the algebraic variable, $\psi$.

The K\"ahler moduli space of the quintic Calabi-Yau threefold is complex one-dimensional and has the structure alluded in section~\Kmodspace. In the Landau-Ginzburg phase of the K\"ahler moduli space the degree five polynomial~\Wquintic\ becomes the superpotential of the Landau-Ginzburg orbifold \WittenYC, where the $\IZ_5$ orbifold group acts on the Landau-Ginzburg chiral fields, $x_\ell$, as
\eqn\QuinticOrbaction{x_\ell\mapsto\omega^k x_\ell \ , \quad \omega\equiv e^{2\pi i\over 5} \ , \quad k\in\IZ_5 \ . }
Thus at the Landau-Ginzburg point of the quintic threefold we adequately represent branes in terms of $\IZ_5$-equivariant matrix factorizations of the quintic polynomial~\Wquintic.

\subsec{Matrix factorization of the quintic threefold}
First we introduce the canonical matrix factorization of the quintic superpotential~\Wquintic. The homogeneous polynomial, $W$, factors as $W={1\over5}\sum_\ell x_\ell\,\partial_\ell W$, which directly yields the canonical matrix factorization, $Q_X$,
\eqn\MFcanQuintic{Q_X\,=\,\sum_{\ell=1}^5\left(x_\ell\pi_\ell+{1\over 5}\partial_\ell W\bar\pi_\ell\right) \ . }
Here $\pi_\ell$ and $\bar\pi_\ell$, $\ell=1,\ldots,5$, are five pairs of boundary fermions, which obey
\eqn\BFCliff{\{\pi_\ell,\bar\pi_k\}\,=\,\delta_{\ell k} \ , \quad \{\pi_\ell,\pi_k\}\,=\,\{\bar\pi_\ell,\bar\pi_k\}\,=\,0 \ . }
These fermions are explicitly realized as a $32\times 32$-matrix representation of this Clifford algebra, and they allow us to express the linear involution, $\sigma_X$, of the canonical matrix factorization as
\eqn\MFcanInv{\sigma_X\,=\, \prod_{\ell=1}^5 \left(\bar\pi_\ell+\pi_\ell\right)\left(\bar\pi_\ell-\pi_\ell\right)  \ . }
The matrix, $\sigma_X$, is the chirality matrix of the Clifford algebra. If we choose a matrix representation for the Clifford algebra~\MFcanQuintic\ such that the involution, $\sigma_X$, is block diagonal, \ie $\sigma_X=\Diag({\bf 1}_{16\times16},-{\bf 1}_{16\times 16})$, then the $32\times 32$ matrix, $Q_X$, decomposes into $16\times 16$ blocks according to eq.~\MFQBlocks. We arrive at a $16\times 16$-matrix factorization in terms of the matrix pair, $(J_X,E_X)$.\foot{Note that also the exceptional factorization~\MFcantorus\ of the cubic torus is the canonical factorization~\MFcanQuintic\ of the homogeneous cubic Landau-Ginzburg superpotential~\Wtorus\ with three boundary fermions \refs{\GovindarajanIM,\GovindarajanUY}.}

The next task is to determine the $U(1)$ R-symmetry representation for the canonical factorization. As the matrix factorization, $Q_X$, and the chiral fields, $x_\ell$, have R~charges $+1$ and $+{2\over 5}$ the boundary fermions, $\pi_\ell$ and $\bar\pi_\ell$, carry R~charges, $+{3\over 5}$ and $-{3\over 5}$, respectively. Therefore along the lines of eq.~\RchargeQ\ the representation, $\rho_X(\theta)$, must act on the boundary fermions as
\eqn\MFRchargeBFaction{
\rho_X(\theta)\,\pi_\ell\,\rho_X^{-1}(\theta)\,=\,e^{3i\theta\over 5} \pi_\ell \ , \quad
\rho_X(\theta)\,\bar\pi_\ell\,\rho_X^{-1}(\theta)\,=\,e^{-{3i\theta\over 5}} \bar\pi_\ell \ . }
Up to an overall phase factor these conditions determine the $U(1)$ R-symmetry representation, $\rho_X$, to be
\eqn\MFXQuinticrho{\rho_X(\theta)\,=\,e^{{3\over 5}i\theta\left(\sum_\ell \pi_\ell \bar\pi_\ell+{\bf 1}_{32\times32}\right)} \ . }
Then with eq.~\OrbRepRcharge\ we readily deduce the five equivariant $\IZ_5$ representations, $R^{X_a}$, for the five branes, $X_a$, $a=1,\ldots,5$, in the equivariant orbit of the canonical factorization~\MFcanQuintic\ 
\eqn\MFQuinticcanequiv{R^{X_a}(k)\,=\,\omega^{ak}\,\omega^{{3\over5}k
\left(\sum_\ell \pi_\ell\bar\pi_\ell+{\bf 1}_{32\times 32}\right)}\,\sigma_X^k \ . }
At the Gepner point in the complex structure moduli space the canonical matrix factorization describes the $L=0$ Recknagel-Schomerus branes \AshokZB. One of these corresponds to the pure D6-brane \BrunnerJQ, and hence also at a generic point in the complex structure moduli space the canonical matrix factorization contains the pure D6-brane in its equivariant orbit.

Next we construct the matrix factorization of the quintic, which contains the D0-brane in its equivariant orbit. Geometrically we describe the locus of the D0-brane as the intersection point of four linear equations, $L_s$, in the ambient projective space, $\IC\IP^4$,
\eqn\QuinticFourLinears{L_s\,=\,\alpha_5 x_s - \alpha_s x_5 \ , \quad s=1,\ldots,4 \ . }
Generically the intersection of these four lines in $\IC\IP^4$ is not located on the hypersurface, $W=0$. If, however, we constrain the parameters, $\alpha_\ell$, to also obey the quintic hypersurface equation
\eqn\AlphaConstQuintic{0\,=\,\sum_{\ell=1}^5 \alpha_\ell^5 - 5 \,\psi\,\alpha_1\alpha_2\alpha_3\alpha_4\alpha_5 \ , }
the intersection point is tuned to lie on the quintic hypersurface. Then the Nullstellensatz ensures that for all parameters, $\alpha_\ell$, fulfilling eq~\AlphaConstQuintic, we can find four quartic polynomials, $F_s$, $s=1,\ldots,4$, such that \BrunnerTC
\eqn\FactRelQuintic{W\,=\,\sum_{s=1}^4 L_s\,F_s \ . }
A view steps of algebra reveal that a possible choice for the quartics, $F_s$, is given by
\eqn\Quartics{F_s\,=\,{1\over\alpha_5^5} \sum_{k=0}^4 (\alpha_s x_5)^{4-k}(\alpha_5x_s)^k
-{5\,\psi\over\alpha_5^s\alpha_s}\left(\prod_{k=1}^s \alpha_k\right)\left(\prod_{k=s}^4 x_{k+1}\right)x_5^{s-1} \ . }

In the final step we use the factorization~\FactRelQuintic\ of the Landau-Ginzburg superpotential to construct again with boundary fermions the matrix factorization, which is associated to the D0-brane at the intersection of the four complex lines~\QuinticFourLinears
\eqn\MFDzeroQuintic{Q_S\,=\,\sum_{s=1}^4 \left( L_s \zeta_s+F_s\bar \zeta_s\right) \ , \quad
\sigma_S\,=\, -\prod_{s=1}^4 \left(\bar\zeta_s+\zeta_s\right)\left(\bar\zeta_s-\zeta_s\right) \ . }
The four pairs of boundary fermions, $\zeta_s$ and $\bar\zeta_s$, $s=1,\ldots,4$, obey the Clifford algebra
\eqn\BFClifftwo{\{\zeta_s,\bar\zeta_t\}\,=\,\delta_{st} \ , \quad \{\zeta_s,\zeta_t\}\,=\,\{\bar\zeta_s,\bar\zeta_t\}\,=\,0 \ , }
and we represent these fermions by $16\times 16$-matrices. Hence choosing a gauge, where $\sigma_S$ becomes $\sigma_S=\Diag({\bf 1}_{8\times 8},-{\bf 1}_{8\times 8})$, we obtain a $8\times 8$-matrix factorization of the matrix pair, $(J_S,E_S)$.

To determine the $U(1)$ R-symmetry representation, $\rho_S$, and the $\IZ_5$-equivariant representation, $R^{S_a}$, we repeat the construction applied to the canonical factorization and we arrive at
\eqn\MFSQuinticrho{\rho_S(\theta)\,=\,e^{{3\over 5}i\theta\left(\sum_s \zeta_s \bar\zeta_s+{\bf 1}_{16\times16}\right)} \ , }
and with eq.~\OrbRepRcharge\ at
\eqn\MFQuinticSequiv{R^{S_a}(k)\,=\,\omega^{ak}\,\omega^{{3\over5}k
\left(\sum_s \zeta_s\bar\zeta_s+{\bf 1}_{16\times 16}\right)}\,\sigma_S^k \ . }

So far we have motivated the matrix factorization, $Q_S$, by geometrically building a D0-brane. The resulting matrix factorization, however, models an orbit of equivariant branes in the non-geometric Landau-Ginzburg phase. Hence it is not obvious that one of the branes, $S_a$, does indeed correspond to the D0-brane. However, by construction the branes, $S_a$, have an open-string modulus parametrized by the parameters, $\alpha_\ell$, which are subject to the constraint~\AlphaConstQuintic. A closer look reveals that the open-string variables, $\alpha_\ell$, are really projective coordinates, because a homogenous rescaling, $\alpha_\ell\rightarrow\lambda\alpha_\ell$, merely generates a gauge transformation~\MFQGauge\ of the factorization, $Q_S$. Hence we observe that the open-string moduli space of the branes, $S_a$, is the quintic threefold, which is the correct open-string moduli space of a D0-brane. In the next section we will present further arguments in favor of this claim.

\subsec{D-brane monodromies on the quintic threefold}
In this section we analyze the monodromies about the singularities in the K\"ahler moduli space of the quintic threefold acting on the matrix factorizations, $Q_X$ and $Q_S$. Since this analysis is similar to the discussion presented in sections~\TorusConMono\ and \TorusMono\ we can be brief here.

We have argued in the previous section that one of the branes, $X_a$, is the D6-brane of the quintic, which we choose to denote by\foot{Since only relative grades \refs{\DouglasGI,\DouglasFJ}, and hence relative equivariant labels, of the branes are physically relevant we are free to choose the D6-brane in the equivariant orbit of the matrix factorization, $Q_X$. This fixes now the grades and equivariant labels of all the other branes.}
\eqn\MFQXquintic{Q_X \equiv Q_{X_1} \ . }
At the conifold point in the K\"ahler moduli space the branes, $X[n]$, become massless \GreeneTX, and hence the factorization, $Q_X$, triggers the transformation~\MFcone\ generated by the monodromy about the conifold point. 

\figinsert\DSixQuiver{The quiver diagram displays the fermionic (red lines) and bosonic (blue lines) open string-states stretching between the D6-brane, $X$, and the other branes, $X_a$, in the same equivariant orbit. The grades of the open-string states are distinguished by different kinds of dashed lines.}{2.2in}{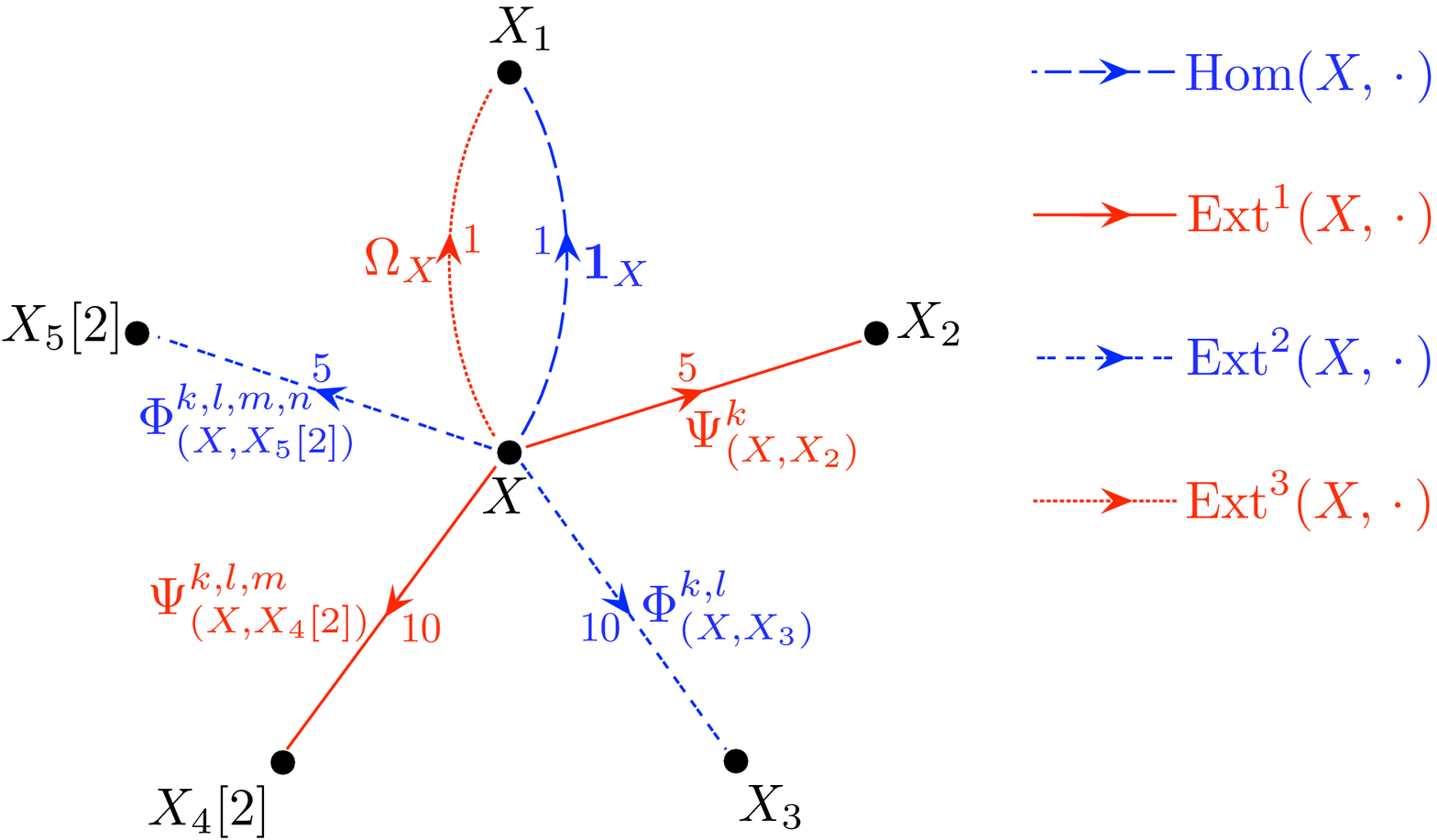}
First we determine the D-brane monodromies associated to the equivariant branes of the canonical matrix factorization, $Q_X$. In order to compute the conifold monodromy we calculate the relevant open-string states stretching between the D6-brane, $X$, and the transported branes, $X_a$. The resulting cohomology elements are summarized in the quiver diagram~\lfig\DSixQuiver.

The bosonic boundary preserving operator, ${\bf 1}_X$, is simply the $32\times32$-identity matrix. The remaining cohomology elements can directly be expressed with the boundary fermions, $\pi_i$ and $\bar\pi_j$. In particular all the open-string states depicted in the quiver diagram are generated by the fermionic open-string states
\eqn\CohXXtwoPsi{\Psi^k_{(X_a,X_{a+1})}\,=\,\pi_k - x_k^3 \bar\pi_k + \psi\,x_{k+1} x_{k+2} x_{k+3}
\bar\pi_{k+4} \ ,\quad k=1,\ldots,5 \ . }
The index of the variable, $x_k$, and the boundary fermion, $\bar\pi_k$, should be thought of taking values modulo $5$. Note that these cohomology elements cannot be exact since the associated matrices carry constant entries arising from the boundary fermion, $\pi_k$. The boundary changing operators~\CohXXtwoPsi\ yield for $a=1$ the five fermionic open-string states, $\Psi^k_{(X,X_2)}$, and give rise to the other states shown in \lfig\DSixQuiver:
\eqn\CohXXList{\eqalign{
\Phi^{k,l}_{(X,X_3)}\,&=\,\Psi^{[k}_{(X,X_2)}\Psi^{l]}_{(X_2,X_3)} \ , \cr
\Psi^{k,l,m}_{(X,X_4[2])}\,&=\,\Psi^{[k}_{(X,X_2)}\Psi^{l}_{(X_2,X_3)}\Psi^{m]}_{(X_3,X_4)} \ , \cr
\Phi^{k,l,m,n}_{(X,X_5[2])}\,&=\,\Psi^{[k}_{(X,X_2)}\Psi^{l}_{(X_2,X_3)}\Psi^{m}_{(X_3,X_4)}\Psi^{n]}_{(X_4,X_5)} \ , \cr
\Omega_X\,&=\,\Psi^1_{(X,X_2)}\cdots\Psi^5_{(X_5,X)} \ .
}}
Here the brackets, $[\ldots]$, indicate that the products are anti-symmetrized.

Now we have assembled all the ingredients to compute the conifold monodromy and to eventually deduce the inverse large radius monodromy of the branes, $X_a$, in the orbit of the canonical matrix factorization. For the conifold monodromy of the brane, $X_1$, we employ again eq.~\MFcone\ and obtain
\eqn\MonoXone{
M_{\rm C}(X_1)\,=\,\pmatrix{
Q_{X_1} & {\bf 1}_X & \Omega_X \cr
0 & Q_X & 0 \cr
0 & 0 & Q_{X[-2]} } \ . }
Due to the thirty-two constant entries arising from the operator, $1_X$, we can remove after a gauge transformation~\MFQGauge\ thirty-two trivial $2\times2$-matrix blocks~\MFtrivial\ and we obtain the simple relation
\eqn\MonoXoneSimple{M_{\rm C}(X_1)\,=\,X_1[-2] \ . }
Thus the conifold monodromy acting on the brane, $X_1$, neither changes its matrix factorization nor modifies its equivariant label, but merely shifts its grade by $-2$. This shift of the D6-brane grade with respect to conifold monodromy has also been observed in ref.~\refs{\AspinwallDZ,\AspinwallJR}, where it was traced back to a simple pole in the period of the D6-brane.

Then we immediately determine the inverse large radius monodromy of the brane, $X_1$, by applying according to eq.~\MFLR\ a subsequent Landau-Ginzburg monodromy 
\eqn\MonoLRXone{M_{\rm LR}^{-1}(X_1)\,=\,X_2[-2] \ . }

In the same fashion we also derive with the open-string states, $\Psi^k_{(X,X_2)}$, the conifold monodromy of the canonical brane, $X_2$, and we find
\eqn\MonoXtwo{
M_{\rm C}(X_2)\,=\,\pmatrix{
Q_{X_2} & \Psi^1_{(X,X_2)} & \cdots & \Psi^5_{(X,X_2)} \cr
0 & Q_X & \cdots & 0 \cr
\vdots & \vdots & \ddots & \vdots \cr 
0 & 0 & \cdots & Q_X } \ , }
whereas the associated $U(1)$ R-symmetry representation and the equivariant representations become
\eqn\MonoXtwoRho{
\rho_{M_{\rm C}(X_2)}\,=\,\Diag\left(\rho_X(\theta),e^{-{2i\theta\over5}}\rho_X(\theta),\ldots,e^{-{2i\theta\over5}}\rho_X(\theta)\right) \ , }
and
\eqn\MonoXtwoEquiv{
R^{M_{\rm C}(X_2)}(k)\,=\,\Diag(R^{X_2}(k),R^{X_1}(k),\cdots,R^{X_1}(k)) \  . }
The presented $96\times 96$-matrix factorization~\MonoXtwo\ is also reducible due to the constant entries in the cohomology elements, $\Psi^k_{(X,X_2)}$. There are a total of $31$ independent constants, which allow us to rewrite the matrix factorization~\MonoXtwo\ to an equivalent $65\times65$-matrix factorization. In this work we do not use and hence do not state the explicit form of the reduced matrix factorization.

The inverse large radius monodromy of the brane, $X_2$, adjusts the equivariant representation of the brane, $M_{\rm C}(X_2)$, 
\eqn\MonoXtwoLREquiv{
R^{M_{\rm LR}^{-1}(X_2)}(k)\,=\,\Diag(R^{X_3}(k),R^{X_2}(k),\cdots,R^{X_2}(k)) \ , }
whereas the matrix factorization, $Q_{M_{\rm LR}^{-1}(X_2)}=Q_{M_{\rm C}(X_2)}$, and the $U(1)$ R-symmetry representation, $\rho_{M_{\rm LR}^{-1}(X_2)}=\rho_{M_{\rm C}(X_2)}$, are not modified.

For the other branes, $X_a$, in the equivariant orbit of the canonical matrix factorization the conifold monodromy and the large radius monodromy are derived analogously. 

\figinsert\DZeroQuiver{The quiver diagram presents the fermionic (red lines) and the bosonic (blue lines) open-string states stretching between the D6-brane, $X$, and the branes, $S_a$. The different blue dashed lines distinguish between the two grades of the bosonic open-string states.}{2.2in}{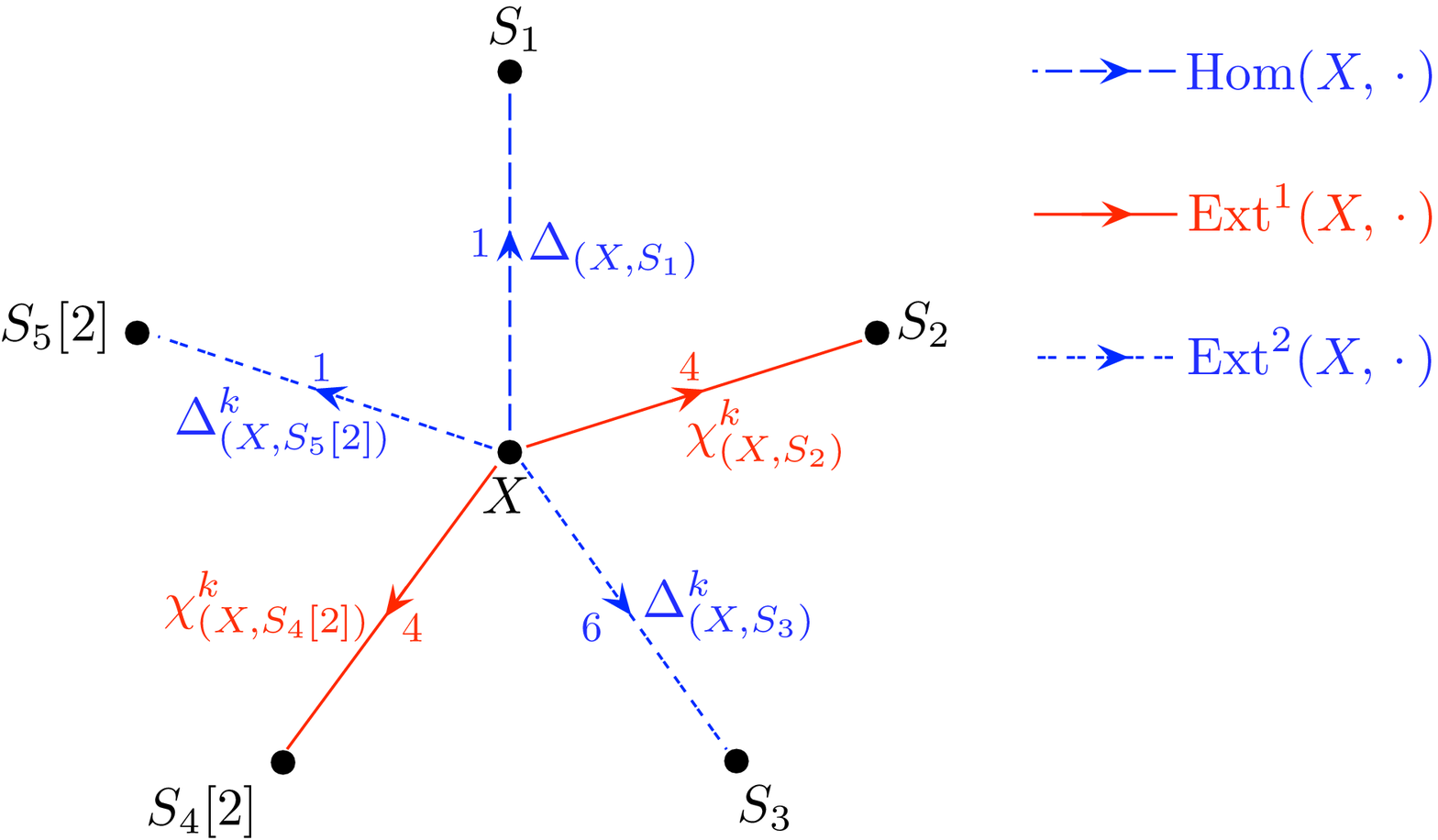}
Finally we want to discuss the monodromies of the branes, $S_a$, in the equivariant orbit of the factorization, $Q_S$. For the monodromy about the conifold point we need to calculate the open-string states between the D6-brane, $X$, and the branes, $S_a$. This is achieved by directly evaluating the cohomology of the BRST~operator~\BRSTOp\  for all possible charge levels~\RchargeTheta\ and equivariant labels~\OrbMFRep. The result of this tedious but straight forward computation is summarized in the quiver diagram~\lfig\DZeroQuiver. We do not present the complicated expressions for the $16\times32$-matrix representation of the open-string states listed in the quiver because for the following analysis we mainly need the multiplicities of the open-string states.

For the brane, $S_1$, the quiver exhibits one bosonic open-string state, $\Delta_{(X,S_1)}$, and hence the monodromy about the conifold point yields with eq.~\MFcone\ the matrix factorization
\eqn\MCQuinticSone{
Q_{M_C(S_1)}\,=\,\pmatrix{Q_{S_1} & \Delta_{(X,S_1)} \cr 0 & Q_{\bar X} } \ . }
The $U(1)$ R-symmetry representation and the $\IZ_5$-equivariant representation become
\eqn\MCQuinticSoneRho{
\rho_{M_C(S_1)}(\theta)\,=\,\Diag\left(\rho_S(\theta), e^{-i\theta}\rho_X(\theta) \right) \ , }
and
\eqn\MCQuinticSoneEquiv{
R^{M_C(S_1)}(k)\,=\,\Diag\left(R^{S_1}(k),R^{\bar X}(k)\right) \ . }
This matrix factorization~\MCQuinticSone\ is again reducible and thus further simplifies with the help of gauge transformations~\MFQGauge\ and by subtracting trivial brane-anti-brane pairs~\MFtrivial. A details analysis reveals
\eqn\MCQuiniticSoneSimple{M_C(S_1)\,=\,S_5[2] \ . }
The shift in the grade and the equivariant label are determined by carefully keeping track of the gauge transformations~\MFQGauge\ acting on the representations~\MCQuinticSoneRho\ and \MCQuinticSoneEquiv. Thus with eq.~\MFequivshift\ we readily deduce for the inverse large radius monodromy
\eqn\MILRQuinticSone{M^{-1}_{\rm LR}(S_1)\,=\,S_1 \ . }

The remaining branes, $S_a$, transform analogously with respect to the monodromies and the analysis is parallel to many previously presented examples. Therefore we immediately turn to the discussion of the RR~charges to gain further insight into the structure of the transformed matrix factorizations. As before we extract the RR~charges of the matrix factorizations by applying the residue formula~\DiskRes. As for the cubic torus,  the residue formula~\DiskRes\ of the quintic hypersurface also reduces to the simplified expression~\DiskResZero. Hence we are able to compute the RR~charges solely from the equivariant representations, $R^{M_C(S_a)}$, which in turn are already determined from the knowledge of the multiplicities of the open-string states depicted in the quivers.

For the quintic hypersurface we obtain non-vanishing disk amplitudes in the twisted sectors, $l=1,\ldots,4$, \ie the potentially non-vanishing correlators with a brane, $P$, are $\braket{1;0}{P}$, \dots, $\braket{4,0}{P}$. Therefore the Landau-Ginzburg charge vector, $\ch_{\rm LG}(P)$, on the quintic threefold is given via eq.~\DiskResZero\ by
\eqn\chLGQuintic{\ch_{\rm LG}(P)\,=\,\left(\braket{1;0}{P},\dots,\braket{4,0}{P}\right) \ . }
The next task is to make the connection to the RR~charges which are natural from a geometric point of view. We denote these charges by the large radius charge vector, $\ch_{\rm LR}$, of the quintic hypersurface
\eqn\chLRQuintic{\ch_{\rm LR}(P)\,=\,\left(d_6,d_4,d_2, d_0\right) \ . }
Here we use the integer basis introduced in ref.~\BrunnerEG, where the integer, $d_6$, denotes the D6-brane charge whereas the lower dimensional brane charges are denoted by $d_4$, $d_2$ and $d_0$.

The Landau-Ginzburg charge vector~\chLGQuintic\ and the large radius charge vector~\chLRQuintic\ are linked with a linear transformation, which we now need to determine. We have argued that the branes, $X_a$, in the orbit of the canonical factorization describe the $L=0$ Recknagel-Schomerus branes, for which on the other hand the large radius RR~charges are recorded in ref.~\BrunnerEG. This allows us to determine the linear transformation we are after. The resulting large radius RR~charges of all the discussed branes are collected in~\ltab\tbRRQuintic.

A closer look at~\ltab\tbRRQuintic\ reveals that the brane, $S_1$, is invariant with respect to the (inverse) large radius monodromy (\cf also eq.~\MILRQuinticSone) and has the large radius charge of a D0-brane. These properties show that the equivariant brane, $S_1$, describes the D0-brane as already anticipated in the previous section.

Finally we observe that the RR~charges of all the branes transform with respect to the (inverse) large radius monodromy as
\eqn\ILRTransQuintic{\otimes\cL^{-1}\ : \ \left(d_6,d_4,d_2,d_0\right)\mapsto
\left(d_6,d_4,d_2,d_0\right) \pmatrix{1 & -1& 5 & -5 \cr 0 & 1 & -5 & 5 \cr 0 & 0 & 1 & -1 \cr 0 & 0 & 0 & 1 } \ . }
This transformation does not change the D6-brane charge, and it turns out that the inverse large radius monodromy acts upon the bundle date of the brane by tensoring with the line bundle, $\cL^{-1}$ \BrunnerEG, where $\cL$ is the line bundle associated to the generator of $H^2({\rm Quintic},\IZ)$. This is the expected transformation behavior associated to the large radius monodromy because physically it corresponds to a shift of the $B$-field by the generator of $H^2({\rm Quintic},\IZ)$. For us the result also serves as a non-trivial check on the computed multiplicities of the open-string states depicted in the quiver diagrams~\lfig\DSixQuiver\ and \lfig\DZeroQuiver.

\tabinsert\tbRRQuintic{For the quintic Calabi-Yau threefold we display the action of the conifold and the large radius monodromy upon the branes, $X_a$ and $S_a$, together with their large radius RR~charges. The RR~charges of the $L=0$ Recknagel-Schomerus branes, $X_a$, have been extracted from ref.~\BrunnerEG\ in order to calibrate the remaining large radius RR~charges.}
{\centerline{\vbox{
\offinterlineskip
\tabskip=0pt\halign{
\vrule height12pt depth6pt#\tabskip=2.5pt plus 1fil\strut &\hfil#\hfil&\vrule# &\hfil#\hfil&\vrule# &\hfil#\hfil&\vrule# 
    &\hfil#\hfil&\vrule# &\hfil#\hfil&\vrule# &\hfil#\hfil &\tabskip=0pt\vrule#\cr
\noalign{\hrule}
&Brane $P_a$ && $\ch_{\rm LR}(P_a)$ && $M_{\rm C}(P_a)$ && $\ch_{\rm LR}(M_{\rm C}(P_a))$
     && $M_{\rm LR}^{-1}(P_a)$ && $\ch_{\rm LR}(M_{\rm LR}^{-1}(P_a))$ &\cr
\noalign{\hrule height1pt}
&$X_1$ && $\ss(1,0,0,0)$    && $X_1[-2]$ && $\ss(1,0,0,0)$   && $X_2[-2]$ && $\ss(1,-1,5,-5)$ &\cr
&$X_2$ && $\ss(1,-1,5,-5)$   && $M_{\rm C}(X_2)$ && $\ss(6,-1,5,-5)$   && $M_{\rm LR}^{-1}(X_2)$ && $\ss(1,-2,15,-20)$ &\cr
&$X_3$ && $\ss(-4,3,-10,5)$   && $M_{\rm C}(X_3)$ && $\ss(-14,3,-10,5)$   && $M_{\rm LR}^{-1}(X_3)$ && $\ss(-4,7,-45,50)$ &\cr
&$X_4$ && $\ss(6,-3,5,0)$   && $M_{\rm C}(X_4)$ && $\ss(16,-3,5,0)$   && $M_{\rm LR}^{-1}(X_4)$    && $\ss(6,-9,50,-50)$ &\cr
&$X_5$ && $\ss(-4,1,0,0)$   && $M_{\rm C}(X_5)$ && $\ss(-9,1,0,0)$   && $M_{\rm LR}^{-1}(X_5)$ && $\ss(-4,5,-25,25)$ &\cr
\noalign{\hrule}
&$S_1$ && $\ss(0,0,0,1)$    && $S_5[2]$    && $\ss(-1,0,0,1)$    && $S_1$   && $\ss(0,0,0,1)$   &\cr
&$S_2$ && $\ss(1,-1,5,-4)$  && $M_{\rm C}(S_2)$  && $\ss(5,-1,5,-4)$   && $M_{\rm LR}^{-1}(S_2)$ && $\ss(1,-2,15,-19)$  &\cr
&$S_3$ && $\ss(-3,2,-5,1)$  && $M_{\rm C}(S_3)$  && $\ss(-9,2,-5,1)$   && $M_{\rm LR}^{-1}(S_3)$ && $\ss(-3,5,-30,31)$  &\cr
&$S_4$ && $\ss(3,-1,0,1)$   && $M_{\rm C}(S_4)$  && $\ss(7,-1,0,1)$   && $M_{\rm LR}^{-1}(S_4)$ && $\ss(3,-4,20,-19)$  &\cr
&$S_5$ && $\ss(-1,0,0,1)$   && $M_{\rm C}(S_5)$  && $\ss(-2,0,0,1)$   && $M_{\rm LR}^{-1}(S_5)$ && $\ss(-1,1,-5,6)$  &\cr
\noalign{\hrule} }}}}

Before we conclude this section we note that, in contrast to the cubic torus, for the quintic hypersurface the large radius monodromy as seen from the gauged linear $\sigma$-model coincides with the large radius monodromy in the Teichm\"uller space. This is due to the fact that the Lefschetz hyperplane theorem ensures that in the gauged linear $\sigma$-model of the quintic the generator of $H^2(\IC\IP^4,\IZ)$ of the ambient projective space, $\IC\IP^4$, induces the generator of $H^2({\rm Quintic},\IZ)$ on the quintic hyperplane \BKGriffithsHarris.

\newsec{Conclusions}
In the context of string compactifications we have probed the global structure of the moduli space by transporting branes along closed loops in the K\"ahler moduli space. Generically the brane probes were transformed along the path as governed by the monodromies of the enclosed moduli space singularities. We chose the base point of these closed loops to be located in the Landau-Ginzburg phase, in which the brane probes were described by matrix factorizations. This required us to develop tools, which were suitable to describe D-brane monodromies  from a matrix-factorization point of view. 

In terms of matrix factorizations the monodromy about the Landau-Ginzburg singularity of the K\"ahler moduli space arose canonically. Following a conjecture of ref.~\KontsevichICM\ we realized the monodromy about the conifold point as a multiple tachyon condensation process of the probe brane with the branes, which became massless at the conifold locus. Finally we computed the action of the large radius monodromy by composing the Landau-Ginzburg and the conifold monodromy. We explicitly demonstrated our techniques on the cubic torus and the quintic Calabi-Yau hypersurface.

A complementary analysis is presented in refs.~\refs{\AspinwallDZ,\DistlerYM,\AspinwallJR}, where the base point for the non-contractible loops is chosen in the large radius regime of the K\"ahler moduli space. In these scenarios D-branes are modeled as complexes of coherent sheaves \refs{\DouglasGI,\LazaroiuJM,\AspinwallPU}, which are then transformed by K\"ahler moduli space monodromies. In this context the conifold monodromy is also realized as a multiple tachyon condensation process. However, the computation of monodromies is rather complicated because generically the probe brane needs to be represented by a suitable complex. Thus in certain situations the computation of the D-brane monodromies is simpler in the language of matrix factorizations as we are able to compute the K\"ahler moduli space monodromies in an algorithmic way.

There are several directions to be further pursued. Our techniques should also apply for hypersurfaces in weighted projective spaces. Furthermore, since matrix factorizations are also a good framework to study obstructed and unobstructed open-string moduli it would be interesting to trace the fate of these moduli with respect to the monodromy transformations alluded here. In this work we have evaded stability issues, which definitely deserve more attention and should eventually be addressed.

\goodbreak
\bigskip
\noindent {\bf Acknowledgments} \medskip \noindent
I would like to thank Manfred Herbst, Nick Warner and especially Wolfgang Lerche for helpful discussions and useful correspondences.

\appendix{A}{Cohomology elements of the `long' and `short' branes}
Here we present explicitly matrix representations of the open-string states displayed in the quiver diagram~\lfig\LSQuiver\ for the `long' and `short' branes of the cubic torus. These matrices are determined by evaluating the BRST cohomology elements~\BRSTOpJE\ depicted in the quiver~\lfig\LSQuiver.

The resulting three fermionic open-string states, $\Psi^k_{(X,L_2)}$, stretching between the D2-brane, $X$, and the `long' brane, $L_2$, are the cohomology elements of the BRST operator, $D_{(X,L_2)}$, and they are given by
\eqn\XLFCohOne{
\Psi_{(X,L_2)}^1 \ : \
\cases{
\psi_0=\pmatrix{
 1 & \left(a-{\alpha_2^2\over\alpha_1\alpha_3}\right)x_1 & -\left(a-{\alpha_1^2\over\alpha_2\alpha_3}\right)x_2 & 0 \cr
 0 & -{\alpha_3\over\alpha_1}x_2 & -{\alpha_1\over\alpha_2}x_3 & -{\alpha_2\over\alpha_3}x_1 \cr
 0 & {\alpha_2\over\alpha_1}x_3 & {\alpha_3\over\alpha_2}x_1 & {\alpha_1\over\alpha_3}x_2} \ , \cr 
\psi_1=\pmatrix{
 \left(a-{\alpha_2^2\over\alpha_1\alpha_3}\right)\alpha_1x_1 & -\alpha_1 & 0 & 0 \cr
 0 & 0 & 0 & -\alpha_3 \cr
-\left(a-{\alpha_1^2\over\alpha_2\alpha_3}\right)\alpha_2x_2 & 0 & -\alpha_2 & 0 } \ , }}
and
\eqn\XLFCohTwo{
\Psi_{(X,L_2)}^2 \ : \
\cases{
\psi_0=\pmatrix{
0 & {\alpha _1\over\alpha_3}x_3 & {\alpha_2\over\alpha_1}x_1 & {\alpha_3\over\alpha_2}x_2 \cr
 1 & 0 & \left(a-{\alpha_2^2\over\alpha_1\alpha_3}\right)x_2 & -\left(a-{\alpha_1^2\over\alpha_2 \alpha_3}\right)x_3 \cr
 0 & -{\alpha_2\over\alpha_3}x_2 & -{\alpha_3\over\alpha_1}x_3 & -{\alpha_1\over\alpha_2}x_1 } \ , \cr 
\psi_1=\pmatrix{
-\left(a-{\alpha_1^2\over\alpha_2\alpha_3}\right)\alpha_2 x_3 & 0 & 0 & -\alpha_2 \cr
\left(a-{\alpha_2^2\over\alpha_1\alpha_3}\right)\alpha_1x_2 & 0 & -\alpha_1 & 0 \cr
0 & -\alpha_3 & 0 & 0} \ , }}
and
\eqn\XLFCohThree{
\Psi_{(X,L_2)}^3 \ : \
\cases{
\psi_0=\pmatrix{
0 & -{\alpha_1\over\alpha_2}x_2 & -{\alpha_2\over\alpha_3}x_3 & -{\alpha_3\over\alpha_1}x_1 \cr
0 & {\alpha_3\over\alpha_2}x_3 & {\alpha_1\over\alpha_3}x_1 & {\alpha_2\over\alpha_1}x_2 \cr
1 & -\left(a-{\alpha_1^2\over\alpha_2\alpha_3}\right)x_1 & 0 & \left(a-{\alpha_2^2\over\alpha_1\alpha_3}\right)x_3} \ , \cr 
\psi_1=\pmatrix{
0 & 0 & -\alpha_3 & 0 \cr
 -\left(a-{\alpha _1^2\over\alpha_2\alpha _3}\right)\alpha_2x_1 & -\alpha_2 & 0 & 0 \cr
 \left(a-{\alpha _2^2\over\alpha_1\alpha_3}\right)\alpha_1x_3 & 0 & 0 & -\alpha_1 } \ . }}
Furthermore evaluating the grading~\OpenGrade\ yields that these fermionic open-string states arise as cohomology elements of $\Ext^1(X,L_2)$.

In the same fashion we deduce the three bosonic open-string states, $\Phi^k_{(X,L_3[2])}$, between the D2-brane, $X$, and the `long' brane, $L_3[2]$. They appear in the  cohomology~\BRSTOpJE\ of the BRST operator, $D_{(X,L_3[2])}$, and turn out to be
\eqn\XLBCohOne{
\Phi_{(X,L_3[2])}^1 \ : \
\cases{
\phi_0=\pmatrix{
\ss -\left(a-{\alpha_2^2\over\alpha_1\alpha_3}\right){\alpha_1\,x_2x_3\over\alpha_2\alpha_3} &\ss {x_1\over\alpha_1} &\ss 0 &\ss 
     -{\alpha_1\,x_2\over\alpha_2\alpha_3} \cr
\ss \left(a-{\alpha_2^2\over\alpha_1\alpha_3}\right){x_3^2\over\alpha_2}
     +\left(a-{\alpha_1^2\over\alpha_2\alpha_3}\right) {\alpha_2\,x_1x_2\over\alpha_1\alpha_3} &\ss 0 &\ss 
     -{\alpha _2\,x_1\over\alpha_1\alpha_3} & {x_3\over\alpha _2} \cr
\ss-\left(a-{\alpha_1^2\over\alpha_2\alpha_3}\right){x_2^2\over\alpha_3} &\ss -{\alpha_3\,x_3\over\alpha_1\alpha_2} &\ss 
      {x_2\over\alpha_3} & 0 } \ , \cr 
\phi_1=\pmatrix{
\ss 1 &\ss 0 &\ss \left(a-{\alpha_1^2\over\alpha_2\alpha_3}\right)x_2 &\ss -\left(a-{\alpha_2^2\over\alpha_1\alpha_3}\right)x_3 \cr
\ss 0 &\ss 0 &\ss 0 &\ss 0 \cr
\ss 0 &\ss {\alpha_1\,x_2\over\alpha_3} &\ss {\alpha_3\,x_3\over\alpha_2} &\ss {\alpha_2\,x_1\over\alpha_1} } \ , }}
and
\eqn\XLBCohTwo{
\Phi_{(X,L_3[2])}^2 \ : \
\cases{
\phi_0=\pmatrix{
\ss \left(a-{\alpha_3^2\over\alpha_1\alpha_2}\right){x_3^2\over\alpha_3}
      +\left(a-{\alpha_2^2\over\alpha_1\alpha_3}\right){\alpha_3\,x_1x_2\over\alpha_1\alpha_2} &\ss
      0 &\ss -{\alpha_3\,x_1\over\alpha_1\alpha_2} &\ss {x_3\over\alpha_3} \cr
\ss -\left(a-{\alpha_2^2\over\alpha_1\alpha_3}\right){x_2^2\over\alpha_1} &\ss -{\alpha_1\,x_3\over\alpha_2\alpha_3} &\ss 
      {x_2\over\alpha_1} &\ss 0 \cr
\ss -\left(a-{\alpha_3^2\over\alpha_1\alpha_2}\right){\alpha_2\,x_2x_3\over\alpha_1\alpha_3} &\ss
      {x_1\over\alpha_2} &\ss 0 &\ss -{\alpha_2\,x_2\over\alpha_1\alpha_3} } \ , \cr 
\phi_1=\pmatrix{
\ss 0 &\ss {\alpha_2\,x_2\over\alpha_1} &\ss {\alpha_1\,x_3\over\alpha_3} &\ss {\alpha_3\,x_1\over\alpha_2} \cr
\ss 1 &\ss 0 &\ss \left(a-{\alpha _2^2\over\alpha_1\alpha_3}\right)x_2 &\ss -\left(a-{\alpha_3^2\over\alpha_1\alpha_2}\right)x_3 \cr
\ss 0 &\ss 0 &\ss 0 &\ss 0 } \ , }}
and
\eqn\XLBCohThree{
\Phi_{(X,L_3[2])}^3 \ : \
\cases{
\phi_0=\pmatrix{
\ss -\left(a-{\alpha_3^2\over\alpha_1\alpha_2}\right){x_2^2\over\alpha_2} &\ss -{\alpha_2\,x_3\over\alpha_1\alpha_3} &\ss 
       {x_2\over\alpha_2} &\ss 0 \cr
\ss -\left(a-{\alpha_1^2\over\alpha_2\alpha_3}\right){\alpha_3\,x_2x_3\over\alpha_1\alpha_2} &\ss
       \ss {x_1\over\alpha_3} &\ss 0 &\ss -{\alpha_3\,x_2\over\alpha_1\alpha_2} \cr
\ss \left(a-{\alpha_1^2\over\alpha_2\alpha_3}\right){x_3^2\over\alpha_1}+
      \left(a-{\alpha_3^2\over\alpha_1\alpha_2}\right){\alpha_1\,x_1x_2\over\alpha_2\alpha_3} &\ss 0 &\ss 
      -{\alpha_1\,x_1\over\alpha_2\alpha_3} &\ss {x_3\over\alpha _1}} \ , \cr 
\phi_1=\pmatrix{
\ss 0 &\ss 0 &\ss 0 &\ss 0 \cr
\ss 0 &\ss {\alpha_3\,x_2\over\alpha_2} &\ss {\alpha_2\,x_3\over\alpha_1} &\ss {\alpha_1\,x_1\over\alpha_3} \cr
\ss 1 &\ss 0 &\ss \left(a-{\alpha_3^2\over\alpha_1\alpha_2}\right)x_2 &\ss -\left(a-{\alpha _1^2\over\alpha_2\alpha_3}\right)x_3} \ . }}
These bosonic open-string states arise with eq.~\OpenGrade\ as cohomology elements of $\Hom(X,L_3[2])$.

Stretching between the D2-brane, $X$, and the `short' brane, $S_2$, we find the fermionic open-string states, $\Psi^k_{(X,S_2)}$, in the cohomology~\BRSTOpJE\ of the BRST~operator, $D_{(X,S_2)}$. These open-string states are elements of $\Ext^1(X,S_2)$ and they read
\eqn\XSFCohOne{
\Psi_{(X,S_2)}^1 \ : \
\cases{
\psi_0=\pmatrix{
\ss 1 &\ss -{a\,x_1\over2}+{\alpha_2\,x_2\over\alpha_3}-\left({3a\over2}-{\alpha_2^2\over\alpha_1\alpha_3}\right){\alpha_1\,x_3\over\alpha_3}
        &\ss -a\,x_2+{\alpha_1^2\,x_3\over\alpha_3^2} &\ss 0 \cr
\ss 0 &\ss {\alpha_1\,x_1\over\alpha_3}-{3a\,x_2\over2}-\left({3a\over2}-{\alpha_1^2\over\alpha_2\alpha_3}\right){\alpha_2\,x_3\over\alpha_3}
        &\ss -{\alpha_2\,x_1\over\alpha_3}+{\alpha_1\,x_2\over\alpha_3}-{\alpha_1\alpha_2\,x_3\over\alpha_3^2} &\ss -x_1 } \ , \cr
\psi_1=\pmatrix{
\ss U_{12} &\ss {x_1\over\alpha_3}-{\alpha_1\,x_3\over\alpha_3^2} 
        &\ss -{\alpha_1^3-\alpha_2^3\over2\alpha_1\alpha_2\alpha_3^2}x_3 &\ss {3a\,x_2\over2\alpha_3}+{x_3\over2\alpha_1} \cr
\ss \alpha_1\,x_1-a\,\alpha_3\,x_2 &\ss 0 &\ss -\alpha_3 &\ss \alpha_2 } \ , }}
and
\eqn\XSFCohTwo{
\Psi_{(X,S_2)}^2 \ : \
\cases{
\psi_0=\pmatrix{
\ss 0 &\ss -{\alpha_2\,x_1\over\alpha_3}+{\alpha_1\,x_2\over\alpha_3}+{\alpha_1\alpha_2\,x_3\over\alpha_3^2}
        &\ss {3a\,x_1\over2}-{\alpha_2\,x_2\over\alpha_3}
                +\left({3a\over2}-{\alpha_2^2\over\alpha_1\alpha_3}\right){\alpha_1\,x_3\over\alpha_3} &\ss x_2 \cr
\ss 1 &\ss a\,x_1-{\alpha_2^2\,x_3\over\alpha_3^2} 
        &\ss -{\alpha_1\,x_1\over\alpha_3}+{a\,x_2\over2}
                 +\left({3a\over 2}-{\alpha_1^2\over\alpha_2\alpha_3}\right){\alpha_2\,x_3\over\alpha_3} &\ss 0} \ , \cr
\psi_1=\pmatrix{
\ss U_{21} &\ss -{\alpha_1^3-\alpha_2^3\over 2\alpha_1\alpha_2\alpha_3^2}x_3 &\ss {x_2\over\alpha_3}+{\alpha_2\,x_3\over\alpha_3^2} 
        &\ss -{x_3\over2\alpha_2}-{3a\,x_1\over2\alpha_3} \cr
\ss a\,\alpha_3\,x_1-\alpha_2\,x_2 &\ss -\alpha_3 &\ss 0 &\ss \alpha_1 } \ . }}
with the quadratic polynomials
\eqn\UPoly{
U_{ij}\,=\, -{a\,x_i^2\over2\alpha_3}-{\alpha_i\,x_j^2\over\alpha_3^2}+{\alpha_j\,x_1x_2\over\alpha_3^2}
    +\left({\alpha_i^3\over\alpha_j}-{\alpha_3^3\over2\alpha_j}+\alpha_j^2\right){x_ix_3\over3\alpha_3^3}
    -\left({\alpha_i^2\over\alpha_j}-{\alpha_j^2\over\alpha_i}\right){a\,x_jx_3\over2\alpha_3^2} \ . }

The bosonic open-string state, $\Phi_{(X,S_1)}$, in $\Hom(X,S_1)$ is they only non-trivial open-string state stretching between the D2-brane, $X$, and the `short' brane, $S_1$,
\eqn\XSBCohOne{
\Phi_{(X,S_1)} \ : \
\cases{
\phi_0=\pmatrix{
{a\,x_1\over2}-{\alpha_2\,x_2\over\alpha_3}+\left({3\,a\over2}-{\alpha_2^2\over\alpha_1\alpha_3}\right){\alpha_1\,x_3\over\alpha_3} 
      & 1 & 0 & -{\alpha_1\over\alpha_3} \cr 
-{\alpha_1\,x_1\over\alpha_3}+{a\,x_2\over2}+\left({3\,a\over2}-{\alpha_2^2\over\alpha_1\alpha_3}\right){\alpha_1\,x_3\over\alpha_3}
      & 0 & -1 & {\alpha_2\over\alpha_3}} \ , \cr 
\phi_1=\pmatrix{
{1\over\alpha_3} & -{a\,x_1\over2\alpha_3}+{\alpha_2\,x_2\over\alpha_3^2}-{x_3\over2\alpha_2} 
      & -{\alpha_1\,x_1\over\alpha_3^2}+{a x_2\over2\alpha_3}+{x_3\over2\alpha_1} 
      & {\alpha_1^3-\alpha_2^3\over2\alpha_1\alpha_2\alpha_3^2}x_3 \cr
0 & \alpha_1 & \alpha_2 & \alpha_3 } \ , }}

Finally between the D2-brane, $X$, and the `short' brane, $S_3[2]$, we find in $\Hom(X,S_3[2])$ the bosonic open-string state, $\Phi_{(X,S_3[2])}$, which reads
\eqn\XSBCohTwo{
\Phi_{(X,S_3[2])} \ : \
\cases{
\phi_0=\pmatrix{
{a\,x_2^2\over\alpha_3}-{\alpha_1\,x_1x_2\over\alpha_3^2} & 0 & {x_2\over\alpha_3} 
     & -{3\,a\,x_1\over2\alpha_3}+{\alpha_2\,x_2\over\alpha_3^2}
         -\left({3\,a\over2}-{\alpha_2^2\over\alpha_1\alpha_3}\right){\alpha_1\,x_3\over\alpha_3^2} \cr
-{a\,x_1^2\over\alpha_3}+{\alpha_2\,x_1x_2\over\alpha_3^2} & {x_1\over\alpha_3} & 0 
     & {\alpha_1\,x_1\over\alpha_3^2}-{3\,a\,x_2\over2\alpha_3}-\left({3\,a\over2}-{\alpha_1^2\over\alpha_2\alpha_3}\right)
        {\alpha_2\,x_3\over\alpha_3^2} } \ , \cr 
\phi_1=\pmatrix{
\left({\alpha_1^2\over2\alpha_2}-{\alpha_2^2\over2\alpha_1}\right){x_3\over\alpha_3^3} &  V_{12} & V_{21} & -{x_1x_2\over\alpha_3^2} \cr
1 & a\,x_1-{\alpha_2\,x_2\over\alpha_3} & {\alpha_1\,x_1\over\alpha_3}-a\,x_2 & 0 } \ , }}
with the quadratic polynomials
\eqn\VPoly{
V_{ij}\,=\,-\left({\alpha_i^2\over\alpha_j}+{\alpha_j^2\over\alpha_i}+{\alpha_3^3\over\alpha_1\alpha_2}\right){x_j^2\over2\alpha_3^3} 
  +{\alpha_i\,x_1x_2\over\alpha_3^3}-\left({\alpha_j^2\over\alpha_i}-{\alpha_i^2\over\alpha_j}\right){a\,x_ix_3\over2\alpha_3^3}
   -{x_jx_3\over2\alpha_i\alpha_3} \ . }

\listrefs
\end